\documentclass[a4paper,11pt]{article}
\usepackage{jinstpub} % for details on the use of the package, please see the JINST-author-manual
\usepackage{lineno}
\usepackage{graphicx}%
\usepackage{multirow}%
\usepackage{amsthm}%
\usepackage{mathrsfs}%
\usepackage[title]{appendix}%
\usepackage{xcolor}%
\usepackage{textcomp}%
\usepackage{manyfoot}%
\usepackage{booktabs}%For nice lines in tables
\usepackage{algorithm}%
\usepackage{algorithmicx}%
\usepackage{algpseudocode}%
\usepackage{listings}%
\usepackage{lineno}
\usepackage[slovak,english]{babel}
\usepackage{caption}
\usepackage{subcaption}

\usepackage[separate-uncertainty=true, free-standing-units=true]{siunitx}
\DeclareSIUnit\eVc{\eV\per\clight\squared}
\DeclareSIUnit\clight{\text{\!\ensuremath{c}}}%add negative spacing '\!' to improve kerning between eV and c^-2

\usepackage{csquotes}
\usepackage{float}
\usepackage{microtype}

\usepackage[version=4]{mhchem}%for chemical formulas like CaWO4
\newcommand{\cawo}{\ce{CaWO_4}}

\usepackage{longtable}%For the long table in appendix
\usepackage{threeparttablex}

\newcolumntype{L}[1]{>{\raggedright\let\newline\\\arraybackslash\hspace{0pt}}m{#1}}
\newcolumntype{C}[1]{>{\centering\let\newline\\\arraybackslash\hspace{0pt}}m{#1}}
\newcolumntype{R}[1]{>{\raggedleft\let\newline\\\arraybackslash\hspace{0pt}}m{#1}}

\usepackage{upgreek}%for upright mu symbol as needed as SI prefix in units

\usepackage{textcomp}%for trademark symbol 'TM'
\usepackage{cleveref}
\crefformat{appendix}{#2#1#3}%To avoid double "appendix Appendix A" see https://tex.stackexchange.com/questions/116675/extra-appendix-when-referencing-using-cleveref-in-elsarticle

\title{\boldmath High-Dimensional Bayesian Likelihood Normalisation for CRESST's Background Model}

\author[a]{G.~Angloher,}
\author[b,c]{S.~Banik,}
\author[d]{G.~Benato,}
\author[a,e]{A.~Bento,}
\author[a]{A.~Bertolini,}
\author[f]{R.~Breier,}
\author[d]{C.~Bucci,}
\author[b,1]{J.~Burkhart,\note{Corresponding author: jens.burkhart@oeaw.ac.at}}
\author[a]{L.~Canonica,}
\author[d]{A.~D'Addabbo,}
\author[d]{S.~Di~Lorenzo,}
\author[b,c]{L.~Einfalt,}
\author[g,h]{A.~Erb,}
\author[g]{F.~v.~Feilitzsch,}
\author[b]{S.~Fichtinger,}
\author[a]{D.~Fuchs,}
\author[a]{A.~Garai,}
\author[b]{V.M.~Ghete,}
\author[d]{P.~Gorla,}
\author[d]{P.V.~Guillaumon,}
\author[b]{S.~Gupta,}
\author[a]{D.~Hauff,}
\author[f]{M.~Ješkovský,}
\author[i]{J.~Jochum,}
\author[g]{M.~Kaznacheeva,}
\author[g]{A.~Kinast,}
\author[b,2]{H.~Kluck,\note{Corresponding author: holger.kluck@oeaw.ac.at}}
\author[j]{H.~Kraus,}
\author[i]{S.~Kuckuk,}
\author[a]{A.~Langenkämper,}
\author[a]{M.~Mancuso,}
\author[d,k]{L.~Marini,}
\author[i]{L.~Meyer,}
\author[b,3]{V.~Mokina,\note{Corresponding author: valentyna.mokina@oeaw.ac.at}}
\author[a]{A.~Nilima,}
\author[d]{M.~Olmi,}
\author[g]{T.~Ortmann,}
\author[d,l]{C.~Pagliarone,}
\author[d,g]{L.~Pattavina,}
\author[a]{F.~Petricca,}
\author[g]{W.~Potzel,}
\author[f]{P.~Povinec,}
\author[a]{F.~Pröbst,}
\author[a]{F.~Pucci,}
\author[b,c]{F.~Reindl,}
\author[g]{J.~Rothe,}
\author[a]{K.~Schäffner,}
\author[b,c]{J.~Schieck,}
\author[b,c]{D.~Schmiedmayer,}
\author[g]{S.~Schönert,}
\author[b,c]{C.~Schwertner,}
\author[a]{M.~Stahlberg,}
\author[a]{L.~Stodolsky,}
\author[i]{C.~Strandhagen,}
\author[g]{R.~Strauss,}
\author[i]{I.~Usherov,}
\author[b]{F.~Wagner,}
\author[g]{M.~Willers,}
\author[a]{V.~Zema,}
\author{(CRESST Collaboration),}
\author[d,m]{F.~Ferella,}
\author[d]{M.~Laubenstein,}
\author[d]{S.~Nisi}

\affiliation[a]{Max-Planck-Institut für Physik, D-80805 München, Germany}
\affiliation[b]{Institut für Hochenergiephysik der Österreichischen Akademie der Wissenschaften, A-1050 Wien, Austria}
\affiliation[c]{Atominstitut, Technische Universität Wien, A-1020 Wien, Austria}
\affiliation[d]{INFN, Laboratori Nazionali del Gran Sasso, I-67100 Assergi, Italy}
\affiliation[e]{LIBPhys-UC, Departamento de Fisica, Universidade de Coimbra, P3004 516 Coimbra, Portugal}
\affiliation[f]{Comenius University, Faculty of Mathematics, Physics and Informatics, 84248 Bratislava, Slovakia}
\affiliation[g]{Physik-Department, Technische Universität München, D-85747 Garching, Germany}
\affiliation[h]{Walther-Meißner-Institut für Tieftemperaturforschung, D-85748 Garching, Germany}
\affiliation[i]{Eberhard-Karls-Universität Tübingen, D-72076 Tübingen, Germany}
\affiliation[j]{Department of Physics, University of Oxford, Oxford OX1 3RH, United Kingdom}
\affiliation[k]{GSSI-Gran Sasso Science Institute, I-67100 L'Aquila, Italy}
\affiliation[l]{Dipartimento di Ingegneria Civile e Meccanica, Universitá degli Studi di Cassino e del Lazio Meridionale, I-03043 Cassino, Italy}
\affiliation[m]{Department of Physical and Chemical Sciences, University of l'Aquila, via Vetoio (COPPITO 1-2), I-67100 L'Aquila, Italy}

\emailAdd{jens.burkhart@oeaw.ac.at}
\emailAdd{holger.kluck@oeaw.ac.at}
\emailAdd{valentyna.mokina@oeaw.ac.at}

\abstract{
Using \cawo{} crystals as cryogenic calorimeters, the CRESST experiment searches for nuclear recoils caused by the scattering of potential Dark Matter particles. A reliable identification of a potential signal crucially depends on an accurate background model. In this work, we introduce an improved normalisation method for CRESST's model of electromagnetic backgrounds, which is an important technical step towards developing a more accurate background model. Spectral templates based on Geant4 simulations are normalised via a Bayesian likelihood fit to experimental background data. Contrary to our previous work, no explicit assumption of partial secular equilibrium is required a priori, which results in a more robust and versatile applicability. %~\cite{CRESST:2022rez}. 
This new method also naturally considers the correlation between all background components. Due to these purely technical improvements, the presented method has the potential to explain up to \SI{82.7}{\percent} of the experimental background within $[\SI{1}{\keV},\SI{40}{\keV}]$, an improvement of at most \SI{18.6}{\percent} compared to our previous method. The actual value is subject to ongoing validations of the included physics. 
}

\keywords{Dark Matter search, Radioactive contamination, Monte Carlo simulation, Geant4, Bayesian inference, Maximum likelihood estimation}

\arxivnumber{2307.12991} % Only if you have one

\begin{document}
\maketitle
\flushbottom

\section{Introduction} \label{intro}
One of the biggest questions in modern physics is the nature of Dark Matter (DM): albeit astronomical observations establish its effects, no particle candidate could be found so far, see e.g.\ \cite{Billard_2022,ParticleDataGroup:2020ssz} for a review.

CRESST is one of the leading experiments that search directly for DM at the sub-\si{\giga\eVc} mass scale at the Laboratori Nazionali del Gran Sasso (LNGS). 
The event signature of a potential DM interaction would be a nuclear recoil inside the target caused by the scattering of a DM particle. 
By default, CRESST operates scintillating \cawo{} target crystals as cryogenic calorimeters, which can measure the phonon and the light signal on an event-by-event basis. 
Combining both signal channels enables the differentiation of nuclear recoils from electromagnetic interactions~\cite{JAGEMANN2006269}. 
Assuming classic leptophobic DM (see e.g.\ \cite{Jungman1996}), the former may be a potential DM signal, whereas the latter is most likely a background event caused by natural radioactivity inside and outside the target crystal.

Recently, CRESST reached detection thresholds for nuclear recoils as low as \SI{10.0}{\eV}, which allows us to constrain spin-independent and spin-dependent DM interactions down to a DM mass below \SI{160}{\mega\eVc}~\cite{CRESST:2022lqw}. 
At such low energy scales, the discrimination power against electromagnetic backgrounds degrades~\cite{Kluck2020} because electromagnetic and nuclear recoil events start to mix due to decreased energy emission in the light channel. 
Hence, a detailed and reliable background description and prediction is crucial. 
Furthermore, it has to be applicable to all of CRESST's detector modules, which are not solely based on \cawo{} crystals, but also other target materials such as Si~\cite{CRESST:2022lqw}, \ce{Al_2O_3}~\cite{Angloher2002}, and \ce{LiAlO_2}~\cite{Angloher2022,CRESST:2022dtl}.

To describe and predict the background, we use a framework that consists of a background \emph{model} (i.e. a set of \emph{physics} assumptions about the background components, and the strength and location of their sources), the \emph{Monte Carlo} (MC) code Geant4~\cite{agostinelli2003geant4,Allison2006,Allison2016} to simulate the background based on the model, and a normalisation \emph{method} for the simulated data (i.e. a set of \emph{technical} rules how to connect the simulated data to the observed data).
The focus of this work is to introduce a new normalisation method to CRESST's background framework and to illustrate how this enables an extension of the existing background model~\cite{CRESST:2019oqe} by considering additional components. The full realisation of this extended background model and a detailed discussion of its impact on CRESST will be the topic of a future publication. Similarly, this study is limited to energies above \SI{1}{\keV} where the principal nature of the observed background (i.e.\ intrinsic contaminations or ambient radiation) is known and the principle applicability of Geant4 is given. The Low Energy Excess of unknown origin at the $\mathcal{O}(\SI{100}{\eV})$ scale~\cite{Fuss:2022fxe,Angloher:2022pas} and the questions how reliable Geant4 simulations are at sub-keV energies~\cite{Kluck:2022lzs} are beyond the scope of this work.

Generally, we normalise the simulation to dedicated screening measurements of the radioactive contaminations and to reference sets of background data recorded with CRESST\@. Based on this, we established the first fully MC-based background model of CRESST in our previous work~\cite{CRESST:2019oqe}, as well as the following procedure, which we call the \textit{Gaussian fit method} (GFM) hereafter: the background activities are determined from parametric Gaussian fits to clearly visible alpha and gamma lines at sideband-measurements and assume partial \textit{secular equilibrium} (SE)~\cite{l2012handbook} to propagate the obtained activity to beta-decaying radionuclides which deposit energy in the energy \textit{range of interest} (ROI), for details see~\cite{CRESST:2019oqe}.

However, the degree to which the partial SE is fulfilled for long-lived radionuclides varies between different crystals due to their production spread~\cite{DANEVICH201144,Danevich_2018} and cannot be known a priori. 
To overcome the assumption of partial SE for future target crystals used by the CRESST experiment, we developed an improved normalisation method based on a combined Bayesian likelihood fit of all considered background components, which we call \textit{likelihood normalisation method} (LNM hereafter).

Besides being a new \emph{normalisation method} with its own advantages, the LNM allows us also to extend the \emph{background model}, i.e.\ we can consider additional background components in our attempt to explain the observed data beyond the ones used in~\cite{CRESST:2019oqe}. In our previous work and in this work, we assume that the background can be entirely explained by electromagnetic interactions caused by bulk contaminations in the crystal and surrounding copper parts; our previous work indicates that surface contaminations are subdominant. The insignificance of neutron background was proven in a recent study~\cite{fuss2022}. In this work, we improve the treatment of several background components: the LNM enables us for the first time to incorporate peak-less spectra (e.g.\ \ce{^3H}) and spectra which share the same peaks but deviate in their continuous part (e.g.\ \ce{^{40}K} in different parts of the setup) directly in the fit. As constraints for the LNM, this work incorporates stricter limits on the contamination in the copper parts of the detectors, which could be set through dedicated screening measurements. We note that future extensions of the background model, e.g.\ including more background components or more precise experimental constraints, may change the predicted background level and composition, but this will not affect the benefits of the LNM that we present here.

Using the TUM40 detector module as a test case, we show the benefits of the LNM (and the improved background model enabled by the LNM) with respect to the previously used GFM (and the original background model limited by the GFM). 
First physics results of the LNM, i.e.\ a study on how strict partial SE is present in a \cawo{} crystal, were already published in \cite{CRESST:2022rez}. 

In \cref{background}, we briefly describe the components of CRESST's background model, including the reference data sets, to provide context for our test case and the benchmark we use for model comparison.
%define our test case and introduce the benchmark we use for the model comparison. 
%We discuss expected improvements of the Likelihood normalisation over the SE normalisation in \cref{ch:likelihood-motivations}. 
Afterwards, we introduce the new LNM in \cref{likelihood-framework}. 
In \cref{copper}, we report the results of the screening measurements, with which we determine some priors for the likelihood. 
We report the outcome of the LNM in \cref{result}, and subsequently, we discuss its performance compared to the previous GFM in \cref{discussion}. 
Finally, we conclude in \cref{summary_outlook}. 
In appendix~\cref{appendixSEUpdate}, we provide the latest results obtained with the GFM as a baseline for comparison, in appendix~\cref{appendixDiffToOld}, we summarise the improvements of the current background model used in this work to the original model developed in \cite{CRESST:2019oqe}, in appendix~\cref{appendixscreening} we give the details of preparation and measurements of the copper samples that lead to the improved contamination limits used in this work, and in appendix~\cref{appendixFullResults} we report the obtained activities of all background components for both normalisation methods: GFM and LNM.

\section{CRESST's Background Model} \label{background}
\begin{figure}[ht]
\centering
%\includegraphics[width=1.0\linewidth]{plots/TUM40.pdf}
%\caption{Visualisation of the TUM40 detector module as implemented in Geant4: \SI{469}{\gram} of copper holds \SI{246}{\gram} of \cawo{} as target.}
\includegraphics[width=0.5\linewidth]{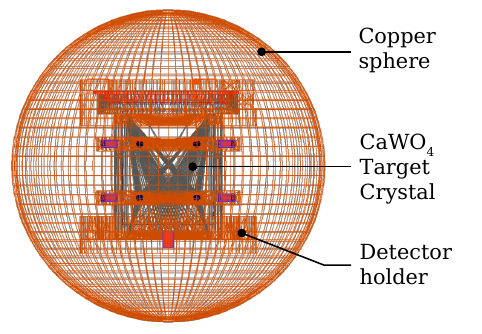}
\caption{Visualisation of the TUM40 detector module as implemented in Geant4: \SI{469}{\gram} of copper holds \SI{246}{\gram} of \cawo{} as target, both are encapsulated by a 1mm-thick copper sphere with a mass of \SI{803}{\gram}.}
\label{fig:tun40geo}
\end{figure}

The GFM was described in our previous work \cite{CRESST:2019oqe}, where we developed a first fully MC-based model for the electromagnetic background of the CRESST experiment using the TUM40 detector module as an example, see \cref{fig:tun40geo} and \cite[fig.\ 6]{CRESST:2019oqe}. To demonstrate the benefits of the LNM over the GFM, we apply the new method also to the TUM40 detector module as our \textit{test case}. Hence, we use a similar set of background sources (\cref{subsec:background-sources}), the same reference data set (\cref{sec:refdata}), and slightly updated spectral templates (\cref{subsec:simulated-background-spectra}). In appendix~\cref{appendixDiffToOld}, we give the detailed differences between the background model used in this work and the previous work.

\subsection{Background Sources}\label{subsec:background-sources}
Our set of physical background sources is split into four background categories: \textit{internal radiogenic} (IR), \textit{internal cosmogenic} (IC), \textit{near external radiogenic} (NER), and \textit{additional external radiogenic} (AER). All categories are treated as bulk contaminations, and no dedicated contribution from surface events is considered.

\begin{figure}[ht]
\centering
\includegraphics[width=0.7\linewidth]{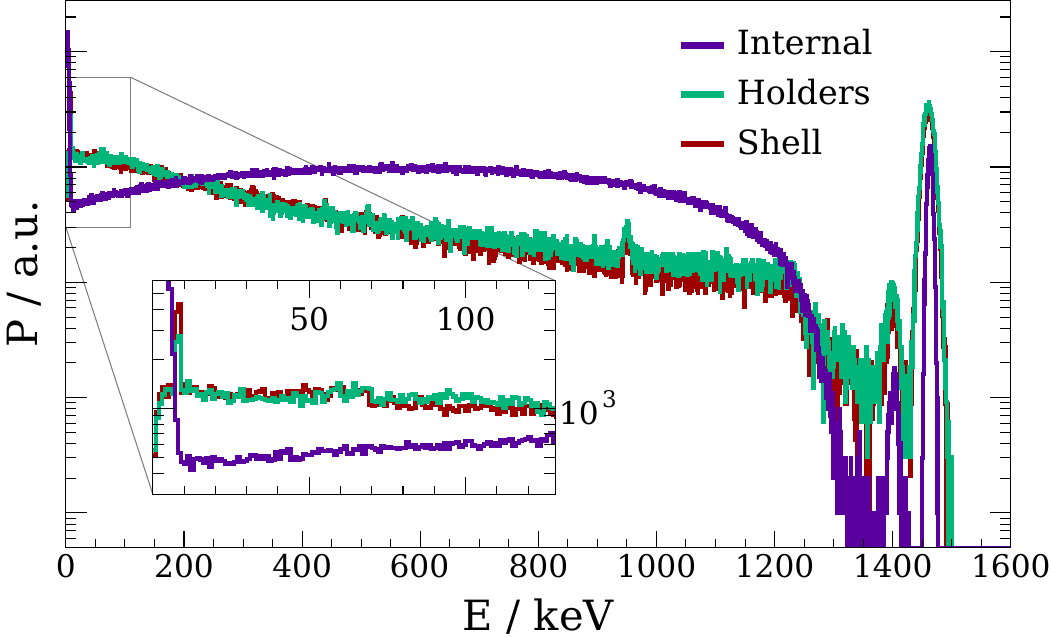}
\caption{Comparison of the simulated spectral templates caused by \ce{^{40}K} in the target crystal (\textit{purple}), in the detector holder (\textit{green}), and in the copper shell (\textit{red}). The \textit{inlay} shows a zoom of the energy range until \SI{120}{keV}. The spectra are normalised to $10^6$ decays and have a bin width of \SI{1}{keV}.}
\label{fig:40K-comparison}
\end{figure}

\begin{figure}[ht]
\centering
\includegraphics[width=0.7\linewidth]{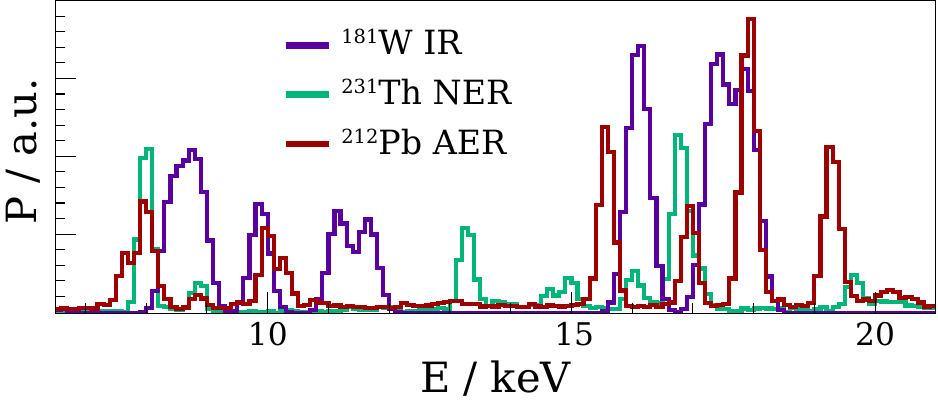}
\caption{Simulated spectral templates caused by \ce{^{181}W} in the target crystal (\textit{purple}), \ce{^{231}Th} in the detector holder (\textit{green}), and \ce{^{212}Pb} in the copper shell (\textit{red}) in a part of the energy range of interest. The normalisation values are arbitrary, and the bin width is \SI{1}{keV}.}
\label{fig:roi-overlapping-peaks}
\end{figure}

In the background model for the LNM, we omit those radionuclides which turned out to be negligible, either because their Q-value lies outside the range of detectable energies or because their branching ratio is insignificant (see \cref{tab:activityPerNuclide} for details). Furthermore, we use the capability of the LNM to discriminate between templates which share the same peaks but deviate in their continuous parts. This allows us to consider e.g.\ \ce{^{40}K} as IR and NER background components (see \cref{fig:40K-comparison}).
See also \cref{fig:roi-overlapping-peaks} as an example where peaks of different nuclides overlap due to the finite detector resolution (see \cref{subsec:simulated-background-spectra}).

For the IR background, we consider the natural contamination of the \cawo{} crystals with \ce{^{40}K} and the 46 radionuclides from the three radioactive decay chains of \ce{^{232}Th}, \ce{^{235}U}, and \ce{^{238}U}. As eleven of these nuclides are dispensable, see \cref{tab:activityPerNuclide}, we consider in total 36 components for IR\@. 
%because they either contribute outside of the detectable energy range or because their expected contribution is negligible (see \cref{tab:activityPerNuclide} for details).
%Additionally, we consider IR \ce{^{40}K} in this work, whereas in the previous model all contributions from \ce{^{40}K} were attributed to the copper sphere source volume.
%This was done because the spectral shapes of \ce{^{40}K} from different source volumes are very similar and the previous normalisation method was not able to disentangle these contributions.
%However, the likelihood normalisation method is able to discriminate between them.
%The background in the ROI is mainly due to the partial absorption of decay radiation from $\beta$ and $\gamma$ decays. Under the assumption of secular equilibrium within the decay chains, the $\beta$ and $\gamma$ activity were previously normalized to the activity of $\alpha$ decays which are clearly visible above the ROI in the high energy range.

The IC background is caused by the activation of the target crystal due to its exposure to cosmic rays during production. 
Only decays of \ce{^3H}, \ce{^{179}Ta}, and \ce{^{181}W} are considered. 
%Via ACTIVIA \cite{BACK2008286} calculations, the \ce{^3H} activity was related to the activity of \ce{^{181}W}.
Due to the capability of the LNM to process featureless, pure beta spectra, the \ce{^3H} activity can be estimated directly from the reference data.
%The activities of \ce{^{179}Ta} and \ce{^{181}W} were previously determined via individual, uncorrelated fits of their clearly visible EC-peaks at $\approx \SI{11}{\keV}$ and $\approx \SI{57}{\keV}$, respectively. Via ACTIVIA \cite{BACK2008286} calculations, the \ce{^3H} activity was related to the activity of \ce{^{181}W}.
We split the contributions of \ce{^{179}Ta} caused by electron capture (EC) into individual templates according to the involved atomic shells because the used Geant4 version (see \ref{subsec:simulated-background-spectra}) wrongly assigns equal capture probabilities regardless of the captured electron's shell of origin. With the capability of the LNM to treat also strongly overlapping peaks, we can disentangle the M-shells. This gives a total number of nine IC components.

The NER background is caused by the radioactive contamination of the detector housing, which is made of NOSV copper (see \cref{copper}) and surrounds the crystal. 
In this work, we consider a subset of 28 out of 46 radionuclides from the three natural decay chains (see \cref{tab:activityPerNuclide}) and assume no SE at all. Together with \ce{^{40}K}, this makes in total 29 components in the NER category.
%that contribute significantly in the ROI\@.
%The contributions of the other 17 nuclides are, again, either outside the detectable energy range or their contribution is negligible due to their low branching ratios. Refer to the footnotes in \cref{tab:activityPerNuclide} for a detailed breakdown.
%Previously, assuming secular equilibrium, their activities were set to values published by CUORE \cite{Alduino_2016} for the same type of copper.

Finally, the AER background consists of ten observed nuclides whose activities cannot be explained entirely by the IR, IC, and NER background components: \ce{^{40}K}, \ce{^{208}Tl}, \ce{^{210}Pb}, \ce{^{212}Pb}, \ce{^{212}Bi}, \ce{^{214}Bi}, \ce{^{226}Ra}, \ce{^{228}Ac}, \ce{^{234}Th}, and \ce{^{228}Ra}. The latter is added in an attempt to explain a yet unidentified peak at \SI{13}{\keV}, see \cref{discussion}. 
%It is assumed that these components originate outside the detector module, in e.g.\ the copper parts of the cryostat or the copper shields of CRESST's setup. As the precise origin is unknown, we approximate the exact geometry of the setup with an arbitrary \SI{1}{\mm}-thick copper sphere around the detector module, see \cref{fig:tun40geo}.
It is assumed that these components originate outside the detector module, in e.g.\ the parts of the cryostat or the shields of CRESST's setup. As the precise origin is unknown, we approximate the geometry of the setup with a copper sphere of an arbitrarily chosen \SI{1}{\mm} thickness around the detector module, see \cref{fig:tun40geo}. We chose copper for the approximation because of all non-modelled components of the CRESST setup, the copper components are those closest to the detector while having a large mass. Hence, we assume them to be the most likely source of the AER background.
%Uncorrelated fits of the remaining peaks were used to determine their activites. It is assumed that this component originates outside the detector module, in e.g.\ the Cu parts of the cryostat or the copper shields of the CRESST setup. As the precise origin is unknown, in \cite{CRESST:2019oqe} we approximate the exact geometry of the setup with a simple copper sphere around the detector module.

%Overall, our background model contains 84 individual background components\footnote{In total, \num{110} components were simulated, but only 84 were used in the final fit. Additional information can be found in \cref{tab:activityPerNuclide}.} in four categories.

\subsection{Reference Data Set}\label{sec:refdata}
For the normalisation, we use data from the detector module TUM40 as experimental \textit{reference data} and the same ROI ($[\SI{1}{\keV},\SI{40}{\keV}]$) as was used in the Dark Matter analysis of TUM40 \cite{angloher2014results}. Due to the peculiarities of CRESST's event reconstruction, the same data are analysed in three different ways, resulting in three non-consecutive data sets $D_j$, which are tuned to and identified by the energy range $E_j$ they cover: \textit{low} $E_\mathrm{l}=[\SI{0.6}{\keV},\SI{495}{\keV}]$, \textit{medium} $E_\mathrm{m}=[\SI{511}{\keV},\SI{2800}{\keV}]$, and \textit{high} $E_\mathrm{h}=[\SI{4}{\MeV},\SI{7}{\MeV}]$ with the \textit{data set index} $j=\{\mathrm{l}, \mathrm{m}, \mathrm{h}\}$. The collected gross exposure is the same for all data sets since target mass and \textit{live-time} $T=\SI{523.8}{\day}$ are identical.

Each set has its own \textit{signal survival probability} (SSP) $\varepsilon_{\mathrm{SSP},j}(E)$ which is independent of the decay process and specific to the experimental setup and the applied event reconstruction, see~\cite{CRESST:2019oqe} for details. 

In the natural decay chains there are three alpha decays with half-lives comparable to the rise time of our signals (below \SI{2}{ms}): $\ce{^{212}Bi} \rightarrow \ce{^{212}Po}$, $\ce{^{214}Bi} \rightarrow \ce{^{214}Po}$, and $\ce{^{219}Rn} \rightarrow \ce{^{215}Po}$. Because of the slow response of the detector, they cannot be resolved in time with respect to the preceding alpha or beta decay. These events are thus seen as one \emph{pile-up} event with the sum of both deposited energies. While those events can be identified in the energy vs.\ light yield plot above \SI{7}{\MeV} (see \cite[fig.4]{strauss2015beta}), their energy can not be reliably reconstructed, and the survival probability of those events can not be determined properly. We, therefore, choose to exclude them from the background model.

\subsection{Simulated Background Spectra}\label{subsec:simulated-background-spectra}
We reuse the simulations of individual background components produced for our previous work with the ImpCRESST physics simulation code~\cite{CRESST:2019oqe} based on Geant4 version 10.2 patch 1. To create \textit{spectral templates} $T_{VXi}$ we apply the following procedure (see~\cite{CRESST:2019oqe} for details): 
(i) in the simulation at time $t=0$, we place $N_{V X,0}$ nuclei of type $X$ in their respective parent volume $V$ (the \cawo{} crystal for IR and IC, the copper detector module for NER, the copper sphere for AER); 
(ii) let them decay but stop the decay after a ground state is reached; this way a decay chain $X_i \rightarrow X_{i+1} \rightarrow X_{i+2} \rightarrow \dots$ is split in subsequent decay steps $X_i \rightarrow X_{i+1}$, $X_{i+1} \rightarrow X_{i+2}$, \ldots which can be individually scaled.
(iii) propagate the decay products to the \cawo{} crystal; 
(iv) record the energy deposited in the target crystal $E_{\mathrm{dep}}$; 
(v) apply the empirical energy and time resolution of the TUM40 detector module to get the observed energy spectrum $E$; compared to our previous work, we apply an improved parameterisation of the
empirical energy resolution $\sigma(E)$, which leads to a better match
between observed and simulated peak widths in our medium
energy data set; 
(vi) count the \textit{simulated entries} $\Tilde{n}_i$ in the $i$-th energy bin $E_i$.

Finally, the spectral template is given as
\begin{equation}
\label{eq:template}
\begin{aligned}
    T_{VXi} &= \frac{\sum_i \Tilde{n}_i(E_i)}{N_{V X,0}} \cdot \frac{\Tilde{n}_i(E_i)}{\sum_i \Tilde{n}_i(E_i)} &= \varepsilon_{\mathrm{g},VX} \cdot f_{VXi}(E_i),
\end{aligned}
\end{equation}
where $f_{VXi}=\Tilde{n}_i(E_i)/\sum_i \Tilde{n}_i(E_i)$ is an empiric \textit{probability density function} (PDF), i.e.\ the probability that an energy deposition in the target by nuclide $X$ in source volume $V$ falls in energy bin $i$, normalised to unity. The \textit{geometric efficiency} $\varepsilon_{\mathrm{g},VX}$ is the ratio of particles that deposit energy in the target over all decays of contaminant $X$ at the source, i.e.\ it accommodates for the cases where there is no energy deposition in the target, e.g. when the decay radiation is absorbed before it can reach the target. 
Together, a spectral template represents the probability that a decay of nuclide $X$ in source volume $V$ causes an energy deposition in the $i$-th energy bin.
We note that the \textit{energy bin index} $i$ depends on the energy range $E_j$ of the data set $D_j$ to which the spectral template should be compared. Also, the actual number of background \textit{templates} is higher than the number of background \textit{components}, as the same component may contribute in several of the three energy ranges $E_j$.

%In total we consider 102 spectral templates.

%In~\cite{turkolu2018development, CRESST:2019oqe} parametric Gaussian templates are used to determine characteristics of clearly distinguishable alpha- and photopeaks. Secular equilibrium assumptions and branching ratios are then used to scale all templates from nuclides within a sub-group of equal activity (see \cref{background}). This alleviates the problem of overlapping beta-spectra and Compton-continua, which would be effectively impossible to disentangle on their own with this method. However, using this normalisation via sideband measurements comes at the expense of consistency and may assume too rigid correlations between activities.

%For the scaling of the decay chains located in the different parts of the detector, the mass of the corresponding part must be taken into account, i.e.\ the activity scales inversely to the mass of the component.

The benchmark is how well the new background model can explain the reference data given the normalisation of its 84 background components\footnote{We note that in \cref{tab:activityPerNuclide}, we list \num{110} components. For the reason why we did not use all of them, see appendix~\cref{appendixFullResults}.}
in four categories (84 free parameters) by the new LNM compared to the old model with its 88 components (30 free parameters) and normalisation by the GFM (see appendix~\cref{appendixSEUpdate}).
%The benchmark is how well the background model can explain the reference data given the normalisation of its 84 background components\footnote{In total, \num{110} spectral templates were simulated, but only 84 were used in the final fit. Additional information can be found in \cref{tab:activityPerNuclide}.} either by the old GFM or the new LNM\@.
%In this context, normalisation means determining the activities $A$ (see \cref{eq:observable-background}) by fitting the spectral templates to the reference data.

\section{Likelihood Framework} \label{likelihood-framework}
% The \emph{likelihood} is defined as the probability of observing the measurement data $\boldsymbol{y}$ given the parameters $\boldsymbol{\vartheta}$. For this reason, it is often interpreted as the \emph{plausibility} of a model. 
Given the problem of normalising multiple spectral templates to make their sum fit the experimental reference data, the LNM is discussed in the following.

\subsection{Motivation for using a Likelihood Normalisation} \label{ch:likelihood-motivations}
The LNM procedure exhibits multiple advantages in comparison to the GFM:
\begin{itemize}
	\item Due to several processing steps during the manufacturing of CRESST's setup components, the SE may be broken within any group of radionuclides that could otherwise be assumed to be in equilibrium. For example, using chemical purification for crystal scintillators can lead to broken SE~\cite{2022, Danevich_2018, DANEVICH201144}. The LNM can overcome the explicit, a priori assumptions of full or partial SE and can recover a degree of SE suitable for the data a posteriori.
	%\item The SE conditions are somewhat arbitrary.
	\item Previously, alpha peaks were fitted and, together with all corresponding beta/gamma-peaks, they were subtracted from the data until all alpha activities were determined. Hence, correlations are only weakly considered since determining the alpha decay rate affects all subsequent rates, but not vice-versa. In the LNM, correlations are inherently considered and are consistently computed as a by-product of the fitting process.
	\item Using the GFM, peak-less
 %featureless or flat
 spectra could previously not be included in the fit on their own without a defined activity value (e.g.\ ${}^{3}\text{H}$ decays). The LNM can directly extract such contamination levels by considering the effects of flat spectra on the fit in the entire energy range.
	\item Spectra with the same peaks or other overlapping features can often be distinguished due to their distinct shapes in their continuous part. See \cref{fig:40K-comparison} for the case of the same contaminant \ce{^{40}K} in different parts of the TUM40 detector module, causing templates with the same photopeak but different continuous parts of the spectra.
%	\item The GFM is lengthy and consists of many manual steps, through which the potential for human mistakes accumulates (see \cref{appendixSEUpdate}).% and~\cite{abdelhameed2019err}).
\end{itemize}
In reconsidering the fitting method of spectral templates, we address and remedy many issues of the GFM by using a powerful, self-consistent, user-friendly, and less time consuming procedure based on Bayesian statistics and the likelihood framework.

It should be noted that this is not the first direct detection DM experiment that uses a likelihood-based method to fit simulated spectral templates as part of their background model. However, this work's fit operates with 84 free floating template normalisation values on a considerably bigger parameter space than previous applications e.g.\ by DAMIC~\cite{aguilararevalo2021characterization} (49 free parameters) and COSINE-100~\cite{govinda2021cosine} ($\lesssim$~40 free parameters).

\subsection{Expectation Value}
\label{sec:expectationValue}
The \textit{expectation value}\footnote{Labeled $N_{\mathrm{obs}}$ in~\cite{turkolu2018development, CRESST:2019oqe}.} $\nu$ of the complete observable background in the $i$-th energy bin of the $j$-th energy range from all contaminants $X$ in all source volumes $V$ is:
%\begin{multline}
\begin{equation}
    \label{eq:observable-background}
    \nu_{ij}(E_i)=\sum_V \sum_X \varepsilon_{\mathrm{SSP},j}(E_i) \cdot \varepsilon_{\mathrm{g},VX} \cdot f_{VXi}(E_i) \cdot T \cdot {A_{VX}} \cdot m_V.
%\end{multline}
\end{equation}
Here, $A_{VX}$ is the \textit{specific source activity} of contaminant $X$ in volume $V$ of mass $m_V$. 

For the sake of simplicity, hereafter we combine the contaminant index $X$ and volume index $V$ to the \textit{process index} $k$ that denotes any process that contributes to the background regardless by which contaminant or from which volume. In this sense, it can e.g.\ also refer to a group of nuclides in SE, which we call \textit{SE groups} hereafter.

Rewriting \cref{eq:observable-background}, we get
%the expected number of events in the $i$-th energy bin of the $j$-th energy range as
~\cite{caldwellBATmanual}
\begin{equation}
	\nu_{ij}(\boldsymbol{\vartheta}) = \sum_{k=1}^{n_\mathrm{p}} \varepsilon_{ijk} \cdot f_{ik} \cdot \vartheta_k,
	\label{eq:expected-events}
\end{equation}
where the process index $k$ runs up to $n_\mathrm{p}=84$ and the data set index $j$ covers the three reference data sets.
Here, $\vartheta_k=T \cdot A_k \cdot m$ is the \textit{normalisation of the process}, i.e.\ the total number of decays in the source volume of process $k$. The \textit{efficiency of the spectral template} $\varepsilon_{ijk}=\varepsilon_{\mathrm{SSP},j}(E_i) \cdot \varepsilon_{\mathrm{g},k}$ is the ratio of observable events over total decay events. The vector of parameters to be marginalised $\boldsymbol{\vartheta}$ consists of the normalisation values $\vartheta_k$.
%total number of decays $\vartheta_k$ of process $k$.
%
Note that the empiric PDF $f_{ik}$ is normalised to the number of simulated events which reach the detector crystal and deposit any energy.

%Note that $f_{ik}$ is the template \emph{after} the energy-dependent influences of geometric attenuation but \emph{before} signal survival efficiency is applied. Since the normalisation of the spectral template $f_{ik}$ equals unity, the factor $\varepsilon_{\mathrm{g},k}$ ensures that the actual number of decay events is used. That is, if a certain number of events is observed in data, one has to compensate for the losses in detection and geometric attenuation to get the number of decaying events. The scaling factor $\vartheta_k$ is then equal to the \emph{total number of decays} of process $k$.

The uncertainty of \cref{eq:expected-events} can be calculated using standard error propagation~\cite{ku1966notes} as was done in~\cite{burkhart2022enhancing}.

\subsection{Likelihood Model}
The $i$-th energy bin in data set $\mathrm{D}_j$ contains a certain \textit{number of measured events} $n_{ij}$ with Poissonian fluctuations that are independent of all other bins.
Using the expected number of detected events $\nu_{ij}$ and the measured number of events $n_{ij}$, one can utilise the Poissonian PDF,
\begin{equation}
	\mathrm{P}(n_{ij}; \nu_{ij}) = \frac{\nu_{ij}^{n_{ij}} \text{e}^{-\nu_{ij}}}{n_{ij}!},
	\label{eq:poisson-pdf}
\end{equation}
%to construct the extended binned log-likelihood\footnote{We note that often in literature, the negative log-likelihood is used because the likelihoods are typically very small, making the log-likelihood negative.} function~\cite{caldwellBATmanual, cowan1998statistical},
%\begin{equation}
%	\log \mathcal{L}(\boldsymbol{\vartheta}) = \sum_{i=1}^{n_{\text{bin}}} (n_i \log \nu_i - \nu_i - \log n_i!),
%	\label{eq:log-likelihood}
%\end{equation}
%which we use for numerical stability instead of the pure likelihood,
%\begin{equation}
%	\mathcal{L}(\boldsymbol{\vartheta}) = \prod_{i=1}^{n_{\text{bin}}} \frac{\nu_{i}^{n_i}\text{e}^{-\nu_i}}{n_i!}.
%	\label{eq:likelihood}
%\end{equation}
to construct the extended binned likelihood\footnote{We note that often in literature, the \textit{negative} log-likelihood is used, allowing the use of popular minimisation tools.} function~\cite{caldwellBATmanual, cowan1998statistical},
\begin{equation}
	\mathcal{L}_j(\boldsymbol{\vartheta}) = \prod_{i=1}^{n_{\text{bin}}} \frac{\nu_{ij}^{n_{ij}}\text{e}^{-\nu_{ij}}}{n_{ij}!}.
	\label{eq:likelihood}
\end{equation}
For numerical stability, we use in the following the extended binned \textit{log}-likelihood function,
\begin{equation}
	\log \mathcal{L}_j(\boldsymbol{\vartheta}) = \sum_{i=1}^{n_{\text{bin}}} (n_{ij} \log \nu_{ij} - \nu_{ij} - \log n_{ij}!).
	\label{eq:log-likelihood}
\end{equation}

The last step in constructing the full likelihood function used in knowledge updating is to combine the likelihoods of the different energy regions ${E}$: low, medium, and high energy reference data (see \cref{background}). The three data sets $\mathrm{D}_\text{l}$, $\mathrm{D}_\text{m}$, and $\mathrm{D}_\text{h}$ can be interpreted as a single data set, making their respective log-likelihoods additive,
\begin{equation}
	\log \mathcal{L}_\text{comb}(\boldsymbol{\vartheta}) = \log \mathcal{L}_\text{l}+\log \mathcal{L}_\text{m}+\log \mathcal{L}_\text{h},
	\label{eq:combined-log-likelihood}
\end{equation}
which is the combined extended binned log-likelihood.

\subsection{Priors and Nuisance Parameters for Bayesian Inference}
%$\log \mathcal{L}_\text{comb}(\boldsymbol{\vartheta})$
\Cref{eq:combined-log-likelihood} is then used as goodness-of-fit (GOF) measure in Bayesian inference of the parameters $\boldsymbol{\vartheta}$. For this, we use an adapted version of the multi-template fitter from the Bayesian Analysis Toolkit (BAT)~\cite{caldwell2009bat}. We use the default settings, i.e.\ the Metropolis-Hastings algorithm (MHA), to efficiently marginalise the posterior parameter distributions. The convergence of the fitting process is discussed in \cref{result}.

%For parameters where activity measurements exist, Gaussian priors are used, otherwise flat priors.
%The latter were restricted by a hard limit in case an upper limit of the activity exists.

Gaussian priors were used for parameters where activity measurements exist.
%Unrestricted flat priors were used for parameters where no previous estimate exists. If activity measurements exist, the priors are Gaussian. 
If an upper limit was determined, the prior is a uniform distribution with an upper boundary. Measured limits of the NER components applied in this work can be found in \cref{tab:reference data}.
Additionally, we use the outcome of the previous background model~\cite{CRESST:2019oqe} as priors for this model.\footnote{The influence on the final fit result was studied and found to be negligible while speeding up the convergence.}
Unrestricted flat priors were used for the AER components, as the GFM and LNM results are found to significantly diverge here, and for \ce{^{179}Ta} (L\textsubscript{3}), where no previous estimate exists (see appendix~\cref{appendixDiffToOld}).

In this work, no marginalised parameter is a nuisance parameter, as we are interested in the activity values of each nuclide. %That is, all nuisance parameters used in the electromagnetic background model are fixed when fitting the $\boldsymbol{\vartheta}$, for example the detector resolution or the reconstructed event energy scale.

\subsection{Implementation and Fit Quality}
\label{subsec:implementation-and-fit-quality}

One marginalisation technique that is widely used today~\cite{robert2004monte} is the generalisation by Hastings~\cite{hastings1970monte} of the Metropolis algorithm~\cite{metropolis1953equation}. It utilises Monte Carlo \emph{Markov chains} (MCMC), which are a sequence of correlated samples such that a random element $X_{n+1}$ depends only on the previous element $X_n$. For each point, a new proposal $X_{n+1}$ is generated according to a \textit{candidate-generating function}, also referred to as \textit{proposal function}. The chain visits regions with high posterior densities more often but is still able to break free to explore new regions with a potentially higher posterior mode. For a more detailed overview of the MHA, we refer to~\cite{brooks2011handbook, chib1995understanding}.

The MHA, as implemented in BAT, is split in two separate phases: the \textit{prerun phase} and the \textit{main run phase}.
During the prerun phase, the scale parameters of the MCMC's proposal function are adjusted. This ensures that the high-dimensional posterior density can be efficiently marginalised. 
In the main run phase, the scaling parameters of the proposal function are fixed, and the MCMC explores the posterior density. % The construction of the MHA ensures that the MCMC visits regions with high density more often, while still being able to break free from local modes.

Note that the MHA is \emph{not} optimised to find a posterior mode but rather to sample the distribution. An optimised algorithm for mode finding is e.g.\ the simulated annealing algorithm (SAA)~\cite{kirkpatrick1983optimization}. It works very similarly to the MHA, with the difference that the acceptance probability is regulated to encourage motion towards the global mode. While the SAA also creates samples of the distribution, it cannot be used for sampling like the MHA since it is strictly not an MCMC technique. Hence, in this work, the MHA is used for sampling marginal distributions while SAA followed by MINUIT~\cite{JAMES1975343} is used for global mode finding.

In practice, not a single but multiple MCMCs run in parallel to ensure that the global mode is sampled.
The chains are initialised at different positions in the parameter space and should converge after a certain number of steps.
Otherwise, they are identical to each other.
In this work's fit, eight chains were used, corresponding to a \textit{high} precision setting of BAT\@.
Using eight chains is a compromise between precision and computational resources.

To monitor the convergence of the fit, we check whether the chains are mixed, i.e.\ explore the same parameter space. This can be done with the $r$-value~\cite{gelman1992inference}. It compares the mean and variance of the expectation value of $\vartheta_k$ for a single chain with the corresponding quantities of multiple chains. For mixed chains, the ratio of these values approaches unity. A commonly used convergence criterion~\cite{caldwellBATmanual} has also been set as $r<1.1$ for this work.

Having assured convergence of the fit, we use multiple well-known statistical similarity measures that quantify the GOF to the experimental data, see~\cite{burkhart2022enhancing}. Additionally, we use two measures that especially helped us by quantifying improvements of the model in an intuitive manner. Firstly, the \emph{reproduction percentage} or \emph {coverage} is defined as the ratio of expected events, as yielded by the normalisation of templates, over the measured number of events in a data set $\mathrm{D}_j$,
%an energy range $\mathbf{E}_j$,
\begin{equation}
	\zeta_j = \frac{N_{\text{MC}}}{N_\text{exp}} \biggr\rvert_j = \frac{\sum_{i=1}^{n_\text{bin}} \nu_{ij}}{\sum_{i=1}^{n_\text{bin}} n_{ij}}.
	\label{eq:coverage}
\end{equation}
It is the primary statistical similarity measure between simulated and experimental data in~\cite{turkolu2018development, CRESST:2019oqe}. Note, however, that~\cref{eq:coverage} does not consider the spectral shape. 

To mitigate this issue and provide a similarly intuitive statistical similarity measure that takes the shape into account, J.\ Burkhart developed the \emph{explainable percentage} ($EP$)~\cite{burkhart2022enhancing}.
It reflects the percentage of events which are well-described by the overall fit and is defined as
\begin{equation}
	EP_j=\frac{\sum_{i=1}^{n_\text{bin}} \Theta\left(p_c(n_i;\nu_i)-\alpha\right)\cdot n_i}{\sum_{i=1}^{n_\text{bin}} n_i},
	\label{eq:explainable-percentage}
\end{equation}
where $\Theta(x)$ is the Heaviside step function, and $p_c(n_i;\nu_i)$ is the central $p$-value~\cite{fay2010two, hirji2005exact} in the $i$-th bin, which is the most-commonly used $p$-value when constructing two-sided hypothesis tests. We assume a \textit{nominal statistical significance} of $\alpha=0.01$, corresponding $\sim 2.58\sigma$.
The Heaviside step function acts as a hypothesis test with the null hypothesis being that the \textit{true} value of the expected number of detected events, $\nu^\text{t}_i$, is indeed given by the predicted value $\nu_i$ from the model, so $\nu^{\text{t}}_i = \nu_{i}$.
If the null hypothesis is rejected, i.e.\ if the central $p$-value is less than the nominal significance level $\alpha$, we take the events in bin $i$ to be \emph{not explainable} by the model.
For more details of this similarity measure, refer to~\cite{burkhart2022enhancing}.

\begin{figure*}
    \centering
    \begin{subfigure}[b]{0.475\textwidth}
    	\centering
    	\includegraphics[width=1.0\linewidth]{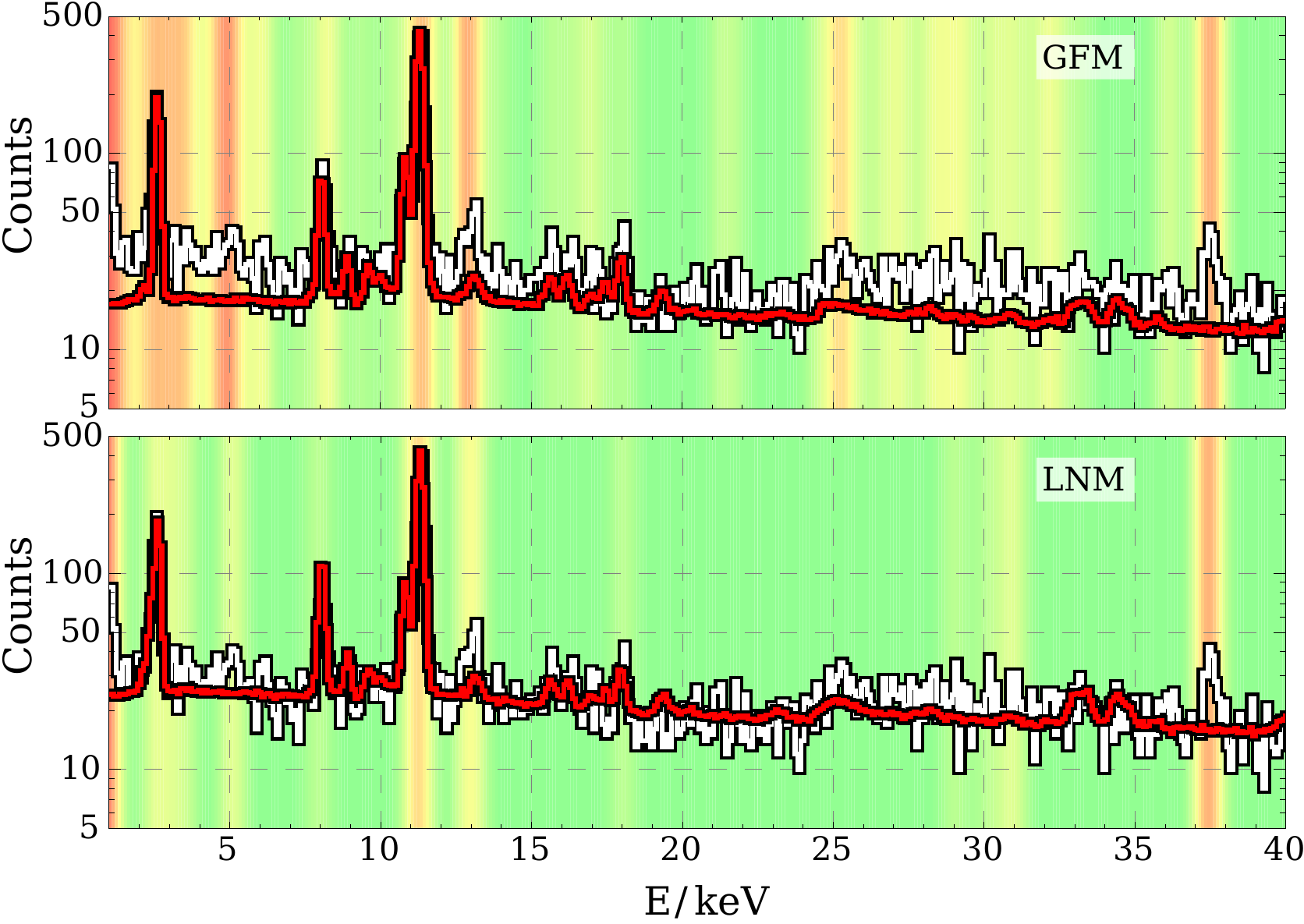}
    	\caption{ROI}
    	\label{fig:hypothesis-plot-comparison-roi}
    \end{subfigure}
    \hfill
    \begin{subfigure}[b]{0.475\textwidth}
    	\centering
    	\includegraphics[width=1.0\linewidth]{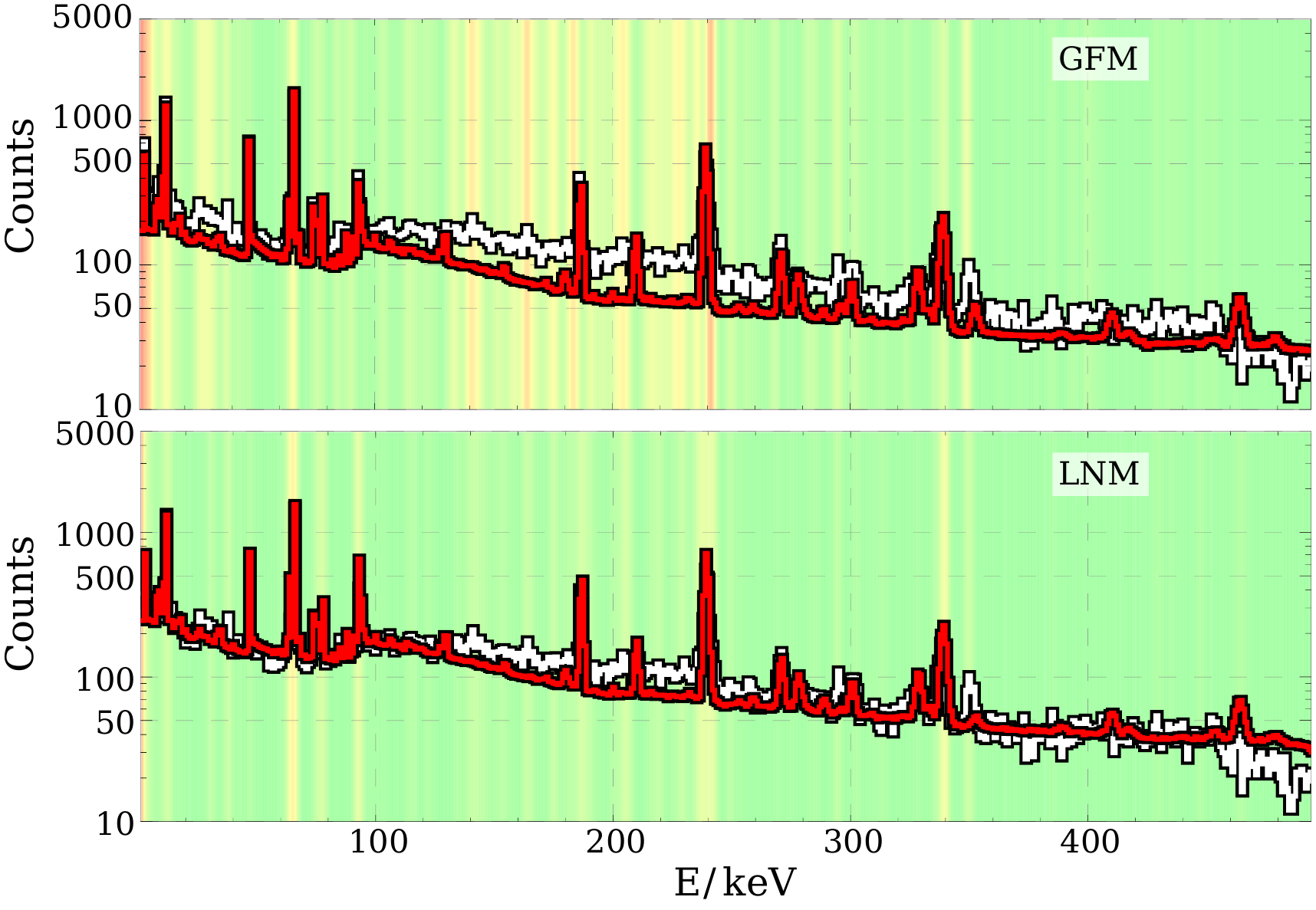}
    	\caption{Low energy range}
    	\label{fig:hypothesis-plot-comparison-low}
    \end{subfigure}
    \vskip\baselineskip
    \begin{subfigure}[b]{0.475\textwidth}
        \centering
        \includegraphics[width=1.0\linewidth]{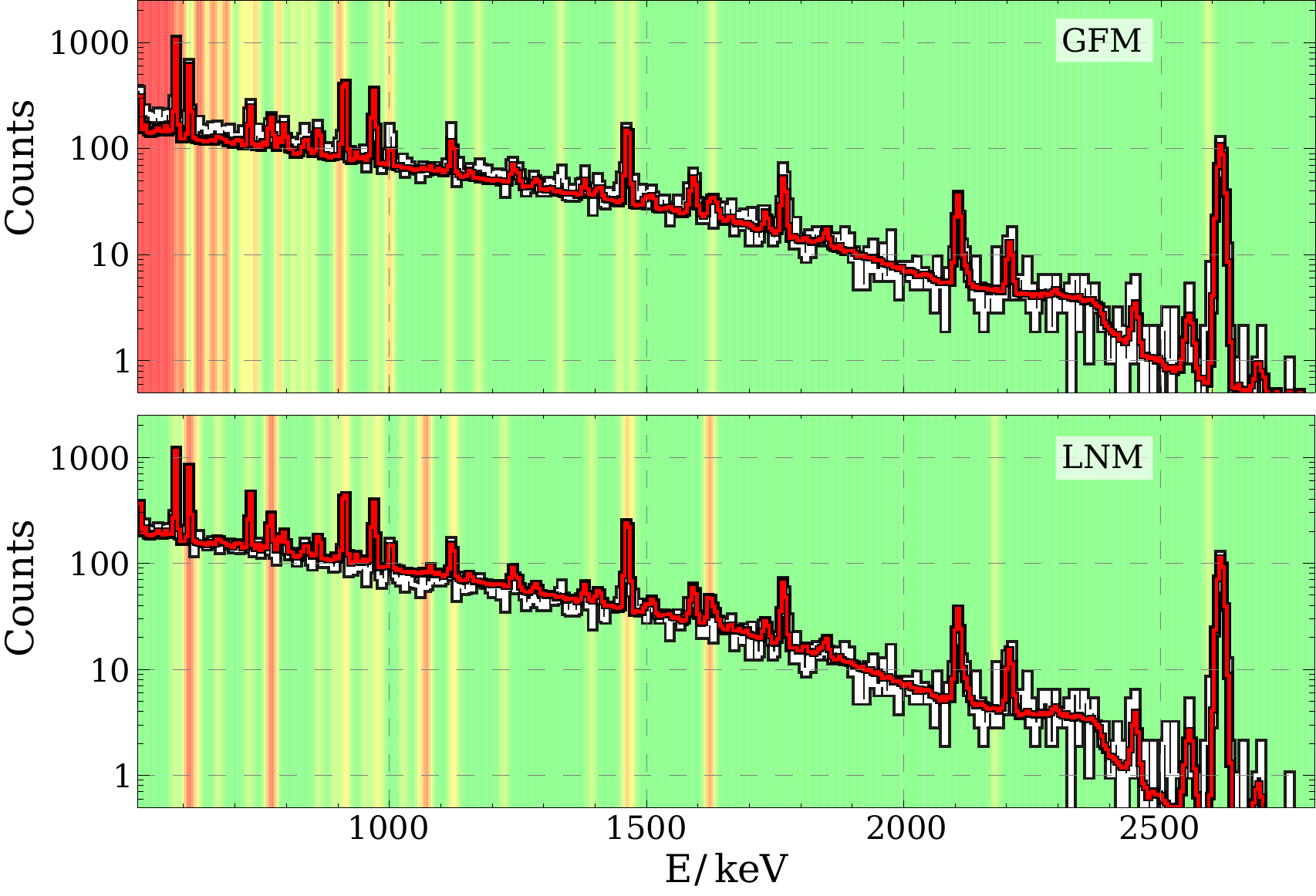}
        \caption{Medium energy range}
        \label{fig:hypothesis-plot-comparison-medium}
    \end{subfigure}
    \hfill
    \begin{subfigure}[b]{0.475\textwidth}
    	\centering
    	\includegraphics[width=1.0\linewidth]{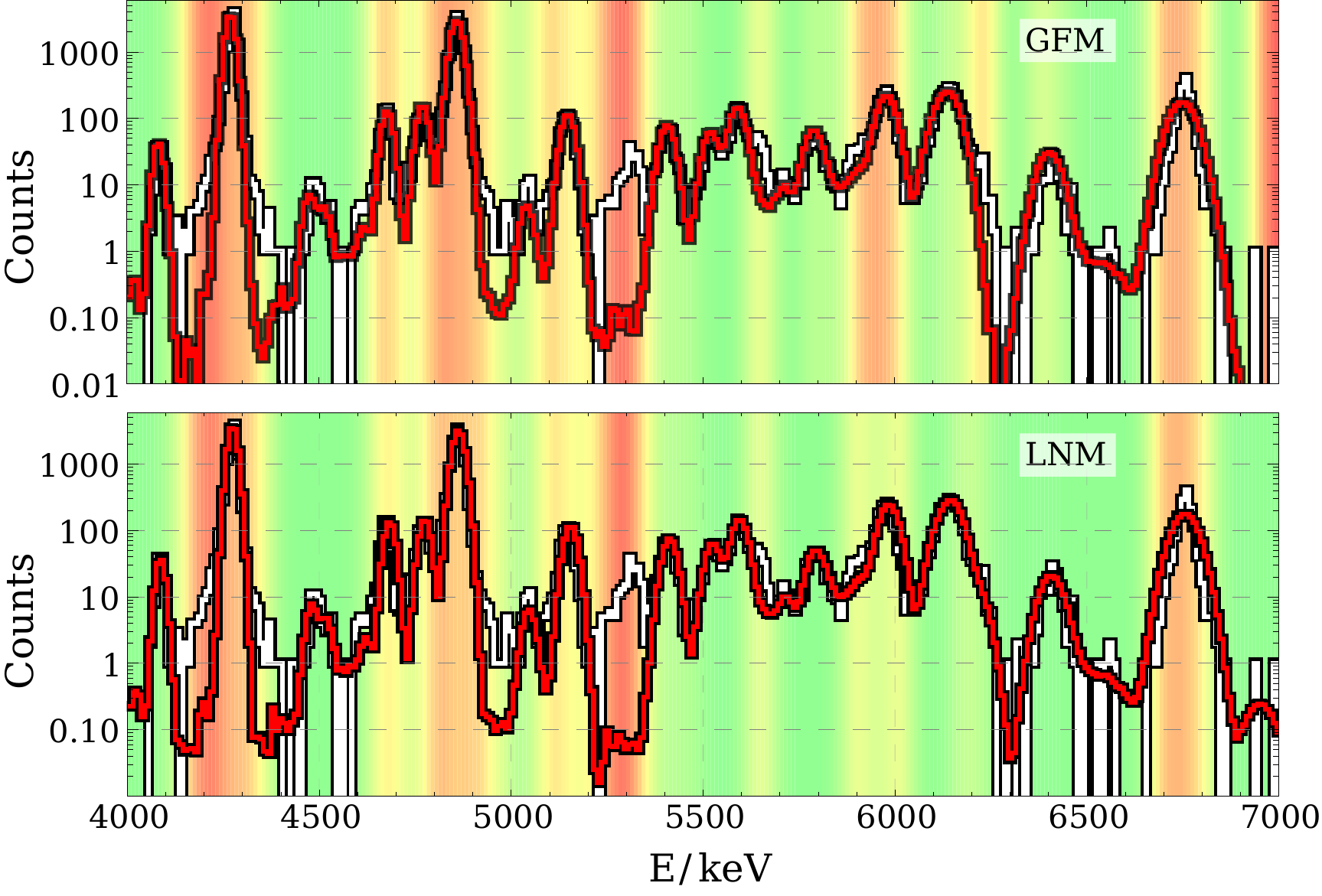}
    	\caption{High energy range}
    	\label{fig:hypothesis-plot-comparison-high}
    \end{subfigure}

    \caption{Hypothesis plots comparing the performance of the updated \emph{Gaussian fit method}~(GFM, see appendix~\cref{appendixSEUpdate}) in the respective \textit{top} panel to the newly developed \emph{likelihood normalisation method} (LNM) in the respective \textit{bottom} panel for (\protect\subref{fig:hypothesis-plot-comparison-roi}) the ROI with \SI{100}{eV} binning, (\protect\subref{fig:hypothesis-plot-comparison-low}) the low energy range with \SI{1}{keV} binning, (\protect\subref{fig:hypothesis-plot-comparison-medium}) the medium energy range with \SI{5}{keV} binning, and (\protect\subref{fig:hypothesis-plot-comparison-high}) the high energy range with \SI{10}{keV} binning. The \textit{white line} represents the experimental data, and the \textit{red line} represents the fit result. The background colour shows the result of hypothesis tests (see \cref{eq:explainable-percentage}) with Weierstrass-smoothened colour values. Each bin is coloured according to the acceptance (\textit{green}) or rejection (\textit{red}) of the null hypothesis that fit and data agree.}
    \label{fig:hypothesis-plots}
\end{figure*}

As another useful application of the $EP$, we will present the \emph{hypothesis plot}~\cite{burkhart2022enhancing} as \cref{fig:hypothesis-plots}.
%in the next section. 
That is, we plot the fit and reference data together with a colour-coded background that summarises the results of the several hypothesis tests that are being conducted in each energy bin $i$ during the evaluation of \cref{eq:explainable-percentage}.
The colour-coding is mainly a visual tool to identify energy ranges where the fit significantly diverges from the data.
The colour values are Weierstrass-smoothened, i.e.\ convolved with a Gaussian, to highlight energy ranges rather than single bins.
In comparison to a conventional residual plot, the hypothesis plot
%(see end of \cref{subsec:implementation-and-fit-quality}) 
contains information about the GOF that is easier to interpret.

\section{Contamination Levels as Prior for Likelihood Normalisation} \label{copper}
\begin{table}[!ht]
    \centering
    \caption{Specific activities of different contaminants measured in NOSV copper samples from CRESST in comparison with limits from CUORE~\cite{alduino2016cuore} used with \textit{Gaussian Fit Method} for the previous work. Limits for high purity germanium detection are at \SI{68}{\percent} C.L., and for inductively coupled plasma-mass spectroscopy detection, they are at \SI{95}{\percent} C.L.}
     \label{tab:reference data}
%\begin{tabular*}{\columnwidth}{@{\extracolsep{\fill}} l c c c}
\begin{tabular}{l S[table-format=5.1(2), separate-uncertainty] S[table-format=3.1] S[table-format=3]}
\toprule\noalign{\smallskip}
\multirow{2}{*}{Component} & \multicolumn{3}{c}{Specific activity / \si{\micro\becquerel\per\kg}}\\
& {HPGe} & {ICP-MS} & {CUORE}\\\midrule
    \ce{^{238}U}            &  {-}          & < 6.2 & < 65\\
    \ce{^{238}U/^{234}Th}   &  < 3500       & {-}   & {-} \\
    \ce{^{238}U/^{234m}Pa}  &  < 760        & {-}   & {-} \\  
    \ce{^{238}U/^{226}Ra}   &  < 20         & {-}   & {-} \\
%    \ce{^{238}U/^{210}Pb}   &  -            &-  &-  \\
    \\
    \ce{^{235}U}            &  < 50         & {-}   & {-} \\
    \\
    \ce{^{232}Th}           &  {-}          & < 2   & < 2\\
    \ce{^{232}Th}/\ce{^{228}Ra}  &  < 24         & {-}   & {-} \\
    \ce{^{232}Th}/\ce{^{228}Th}  &  < 20         & {-}   & {-} \\
    \\
    \ce{^{137}Cs}           &  < 5.6        & {-}   & {-} \\
    \ce{^{40}K}             &  < 190        & {-}   & {-} \\
    \\
    \ce{^{60}Co}            & 46 +- 6       & {-}   & {-} \\
    \ce{^{59}Fe}            & 42 +- 11      & {-}   & {-} \\
    \ce{^{58}Co}            & 500 +- 50     & {-}   & {-} \\
    \ce{^{57}Co}            &  < 140        & {-}   & {-} \\
    \ce{^{56}Co}            & 54 +- 8       & {-}   & {-} \\
    \ce{^{54}Mn}            & 51 +- 9       & {-}   & {-} \\
    \ce{^{48}V}            & < 40          & {-}   & {-} \\
    \ce{^{46}Sc}            & 29 +- 6       & {-}   & {-} \\
   \noalign{\smallskip}\bottomrule
    \end{tabular}
\end{table}

%\subsection{Activity values obtained from experimental reference data sets} \label{subsec:previous_activities}
%Since use the same experimental \textit{reference data} as in~\cite{CRESST:2019oqe}, we used as prior Gaussian fit values of gamma and alpha lines observed in these data sets. Taking to account that for previous work we used activity limits for copper from CUORE experiment~\cite{Alduino:2016vjd}, we organized our own screening measurements which are described bellow. 
%Due to the lack of the needed screening results, as priors for the likelihood normalisation we use contamination %levels obtained in our previous work~\cite{CRESST:2019oqe} and from literature.
For the IR, IC, and AER background components, we use contamination levels obtained in our previous work~\cite{CRESST:2019oqe} via GFM from the reference data 
%and from literature
as priors for the likelihood normalisation.
This is due to the absence of activity values or limits from dedicated screening measurements.
For the contaminants that cause the NER, we previously used activity limits for the heads of natural decay chains for copper from the CUORE experiment~\cite{alduino2016cuore}, which we assumed to be from the same batch. For this work, we use limits for the same radionuclides and \ce{^{40}K} (in addition) obtained from our own measurements during a screening campaign, which was initiated to define the level of radiopurity in different materials used and planned to be used by the CRESST experiment.
%The values obtained for radionuclides from cosmogenic activation are planned to be included in a future extended background model. 
%\subsection{Copper sample preparation} \label{subsec:screening}
%\subsubsection{Sample preparation} \label{subsubsec:sample}

The NER background component originates from a copper housing produced from electrolytic tough pitch type copper, also known as NOSV, %\footnote{Aurubis AG, \url{https://www.aurubis.com/en/products/copper-products}}, 
which has \SI{99.9975}{\percent} purity and is suitable for cryogenic use. NOSV copper is extensively used in the experiment for the construction of detector modules, the detectors' holding structure, and components of the cryogenic infrastructure.

To determine the radiopurity of a copper sample, we used inductively coupled plasma-mass spectrometry (ICP-MS) and ga\-mma-ray spectrometry using a high-purity germanium (HPGe) detector. The samples's preparation procedure for both methods, including commonly known surface cleaning treatments, is described in appendix \cref{subsec:screening}. 
A short description of the measurement using an HPGe detector is presented in appendix \cref{subsubsec:hpge}. The details of the ICP-MS measurement are given in appendix \cref{subsubsec:icp-ms}. The results of both are listed in \cref{tab:reference data}. The values obtained for radionuclides from cosmogenic activation (\ce{^{46}Sc} to \ce{^{60}Co}) and for \ce{^{137}Cs} are planned to be included in a future extended background model. 

In summary, the limit of \ce{^{238}U} achieved in this work is more sensitive compared to the one from CUORE~\cite{alduino2016cuore}, which shows that in the previous work~\cite{CRESST:2019oqe} the contribution of NER was overestimated.

%The list of internal radioactive nuclides are listed in \cref{tab:reference data}.

\section{Outcome of the Fit Process} \label{result}
%The methods presented in the previous \cref{subsec:implementation-and-fit-quality} were combined into a program that executes the likelihood fit. After its successful completion, the results from the program are analysed in this section. The convergence of the MCMC is checked first. % After that, the goodness of fit is determined.

%\subsection{Assessment of Fit Convergence and Quality}\label{sec:assessment-of-fit-convergence-and-quality}

Using the activity values from~\cite{CRESST:2019oqe} as priors for the IR, IC, and AER components, and the lowest limit for each nuclide from \cref{tab:reference data} for the NER component, we executed the Bayesian likelihood normalisation on the Vienna computing cluster CLIP\@. A typical, parallelised run takes less than 24 CPU hours.

\begin{figure}[ht]
	\centering
	\includegraphics[width=0.6\linewidth]{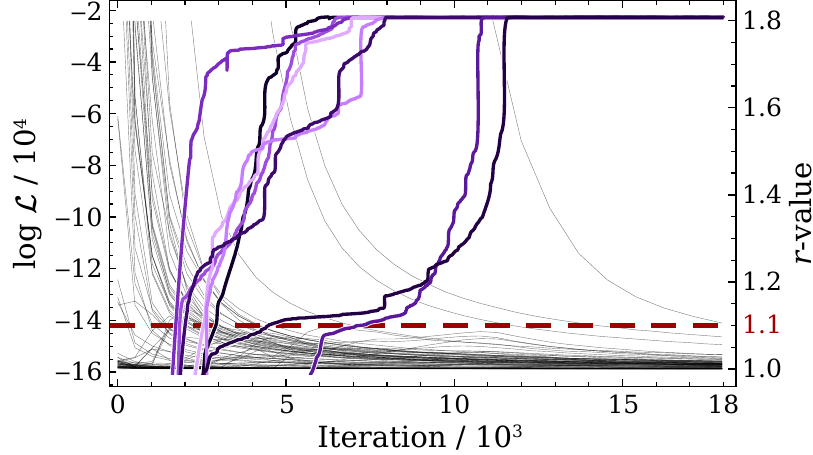}
	\caption{Log-likelihood during the prerun phase for all eight Monte Carlo Markov chains (\textit{purple} solid lines; \textit{left y-axis}). The chains are identical but start at different positions in the parameter space to make the fitting procedure more robust against getting stuck in local modes. The $r$-value of all 84 parameters are shown as \textit{black} lines, the convergence threshold value of \num{1.1} is drawn as \textit{dashed, red} line (\textit{right y-axis}).}
	\label{fig:noSE-newLims-convergence-ll}
\end{figure}

\Cref{fig:noSE-newLims-convergence-ll} shows the convergence  of all eight MCMCs during \num{18000} prerun iterations (see \cref{subsec:implementation-and-fit-quality}). Using multiple Markov chains %\footnote{Note that despite the similar name, Markov \textit{chains} have \textit{no} relation to radioactive decay \textit{chains}.} 
ensures a more robust marginalisation and reduces the risk of getting stuck in a local mode.
%Note that MCMC chains are different from radioactive decay chains.

\begin{figure}[ht]
	\centering
	\includegraphics[width=0.5\linewidth]{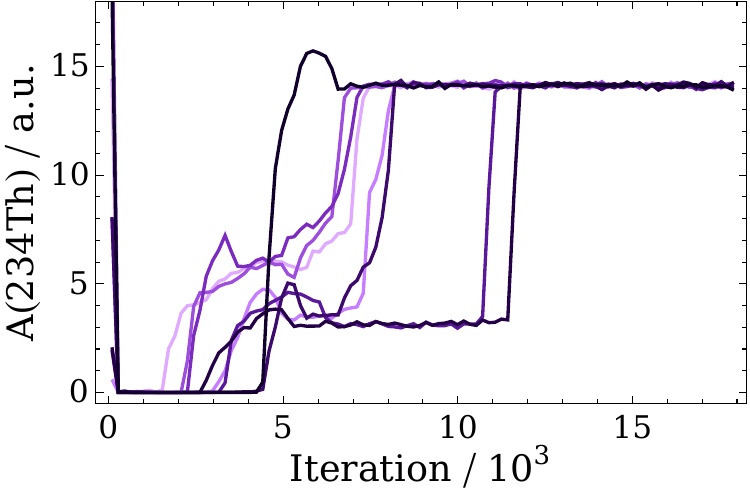}
	\caption{The evolutions of the IR ${}^{234}\text{Th}$ activity value in arbitrary units during the prerun phase for all eight Monte Carlo Markov chains (\textit{purple} lines). The chains are identical but start at different positions in the parameter space to make the fitting procedure more robust against getting stuck in local modes.
    %Note how some of the chains are temporarily stuck in local modes before converging to a common value.
    Thanks to the Metropolis-Hastings algorithm, temporarily stuck chains can break out of local modes and finally converge to a common value.}
	\label{fig:noSE-newLims-convergence-234th}
\end{figure}

%Of the 110 simulated spectral templates, only 83 have a non-negligible contribution in the fitting range, see \cref{tab:activityPerNuclide}. 
%As an example\footnote{Note that this marginalisation is done for all \num{84} parameters in parallel.}, \cref{fig:noSE-newLims-convergence-234th} shows the evolutions of $\vartheta_\mathrm{234Th}$, the expected activity value of IR ${}^{234}\text{Th}$ decays.
All \num{84} parameters are marginalised in parallel; as an example, \cref{fig:noSE-newLims-convergence-234th} shows the evolution of the expected activity value $A_\mathrm{234Th}$ of IR ${}^{234}\text{Th}$ decays.
After \num{18000} iterations in the prerun, the chains appear to be mixed, considering the agreement between all chains within each parameter. 

This conclusion is further supported by the fact that the $r$-values of all \num{84} parameters, which represent the activities of 84 decay processes, fall below the limit of \num{1.1} within the adjustment phase, see \cref{fig:noSE-newLims-convergence-ll}.

\begin{figure}[ht]
	\centering
	\includegraphics[width=0.52\linewidth]{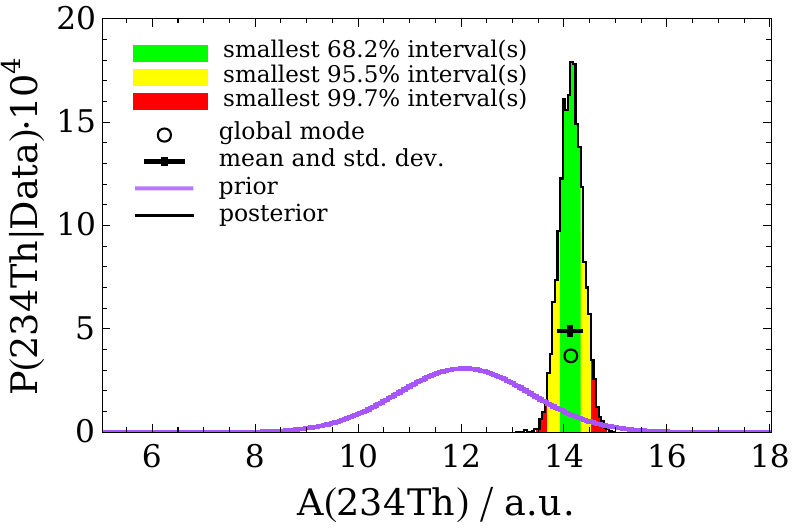}
	\caption{The 1D marginalisation of the activity of IR ${}^{234}\text{Th}$ (\textit{coloured} histogram) in arbitrary units, together with its prior (\textit{purple} histogram). See also \cite{CRESST:2022rez} for a comparison of priors and posteriors.}
	\label{fig:noSE-newLims-convergence-234th-marginalised}
\end{figure}

The main run phase now marginalises the posterior distribution for \num{2500} iterations to determine the parameters' mean values, uncertainties, and correlations. \Cref{fig:noSE-newLims-convergence-234th-marginalised} depicts an exemplary marginalised distribution together with its prior. As can be seen there, the marginalised posterior distribution of the IR parameter $A_\mathrm{234Th}$ agrees well with the value found in~\cite{CRESST:2019oqe}, i.e.\ the prior. However, the uncertainty is considerably smaller as can also be seen in \cref{tab:activityPerNuclide}.

%\begin{figure}[ht]
%	\centering
%	\includegraphics[width=1.0\linewidth]{plots/r_value}
%	\caption{$r$-values (see \cref{subsec:implementation-and-fit-quality}) during the prerun phase for all \num{84} parameters. The %\textit{red line} indicates the upper convergence limit of \num{1.1}.}
%	\label{fig:noSE-newLims-r-value}
%\end{figure}

\section{Discussion of Fit Results} \label{discussion}
\label{subsec:goodness-of-fit-results}\label{subsec:component-decomposition}

%The GOF of the LNM improved relative to the GFM: for example, the Kolmogorov-Smirnov (KS) divergence drops from \num{0.029} to \num{0.024} in the ROI\@. This means that a two-sample KS test, i.e.\ whether the combined fit spectrum and the experimental background data have the same underlying PDF, can be rejected at significance levels down to \num{1e-7} and \num{3e-5} ($\num{5.2}\sigma$ and $\num{4.0}\sigma$), respectively. In other words, even though in both cases the evidence for a different underlying probability distribution is high (the null hypothesis is rejected for very small $\alpha$), the test rejects the GFM case more strongly. To avoid systematic biases, we evaluate also other GOF metrics, see~\cite{burkhart2022enhancing} for the results; all of them confirm the improvement. Overall, this LNM-based work can reproduce $\zeta_\mathrm{ROI}=\SI{93.9}{\percent}$ of the background observed in the ROI of TUM40\@.

Overall, this LNM-based work has the potential to attribute significantly more background than the work based on GFM: measured as an explainable percentage $EP$ (\cref{eq:explainable-percentage}), the increase in the ROI may go from \SI{64.5}{\percent} for the GFM to \SI{82.7}{\percent} for the LNM, see \cref{tab:comparison-explainable-percentage-cenk-noSE-newLims}. Measured as coverage $\zeta$ (\cref{eq:coverage}), this would be an improvement in the ROI from \SI{74.6}{\percent} to \SI{93.9}{\percent}, see \cref{tab:comparison-coverage-cenk-noSE-newLims}.

This improvement is also supported by e.g.\ the Kolmogo\-rov-Smirnov (KS) divergence as GOF: it drops from \num{0.029} for the GFM to \num{0.024} for the LNM in the ROI\@. This means that a two-sample KS test, i.e.\ whether the combined fit spectrum and the experimental background data have the same underlying PDF, can be rejected at significance levels down to \num{1e-7} ($\num{5.2}\sigma$) for the GFM but only down to \num{3e-5} ($\num{4.0}\sigma$) for the LNM: even though in both cases the evidence for a different underlying probability distribution is high (i.e.\ the fitted spectrum misses some observed background components), the test rejects the LNM less strongly (i.e.\ the LNM can fit the background better). To avoid systematic biases, we evaluate also other GOF metrics, see~\cite{burkhart2022enhancing} for the results; all of them confirm this improvement.

To demonstrate the technical capabilities of the LNM, in the following, we will present its outcome with respect to the GFM in terms of: the contribution of individual contaminants from the four background categories IR, IC,
NER, and AER to the combined background spectrum; the power of the LNM to handle correlated contaminants; the improved overall spectral agreement; and the impact of omitting the SE assumption on the results.

For the convenience of the reader, we first introduce the relevant figures and tables to which we will frequently refer: 
\begin{itemize}
    \item The activity values $A$ of each background component as obtained with both methods are listed in \cref{tab:activityPerNuclide}.
    \item \Cref{fig:noSE-newLims-roi-stacked-plot,fig:noSE-newLims-low-stacked-plot,fig:noSE-newLims-medium-stacked-plot,fig:noSE-newLims-high-stacked-plot} show the related LNM background components in comparison to the reference data in the respective energy range. 
    \item The coverage $\zeta$ (\cref{eq:coverage}) for each energy range is listed in \cref{tab:comparison-coverage-cenk-noSE-newLims} and its decomposition into the four background categories is given in \cref{tab:decomposition} and shown in \cref{fig:decompositions} in comparison with GFM results.
    \item The statistical agreement with the experimental reference data in terms of explainable percentages $EP$ (\cref{eq:explainable-percentage}) is listed in \cref{tab:comparison-explainable-percentage-cenk-noSE-newLims} and shown in the hypothesis plots \cref{fig:hypothesis-plots} (see also \cref{subsec:implementation-and-fit-quality}).
\end{itemize}
%We discuss -- for all energy ranges -- the decomposition of the coverage for each of the four background categories IR, IC, NER, and AER, and the agreement of spectral features.

\subsection{Alpha Background}

The spectrum in the high energy range consists mainly of clearly visible peaks of alpha-decaying nuclides. Due to their short interaction lengths, these alphas stay inside the crystal and belong entirely to the IR background. 
The feature-rich data are well usable for both GFM and LNM. Hence the coverage improved only by \SI{2.0}{\percent} (\cref{tab:comparison-coverage-cenk-noSE-newLims}) to \SI{100.2}{\percent}~\footnote{This overestimation is of no concern, as it is only \emph{local} to the high energy data set; the \emph{global} coverage of all three data sets with \SI{99.6}{\percent} does not overestimate the background.}.
Compared to the GFM, the LNM is able to improve the spectral agreement by \SI{7.2}{\percent} (\cref{tab:comparison-explainable-percentage-cenk-noSE-newLims}) in terms of $EP$. 

%With our current study, we significantly improve the agreement between simulation and data in the energy range between the \ce{^{234}U} peak at \SI{4.857}{\MeV}\footnote{The literature data for energy levels, half-lives $T_\mathrm{1/2}$, and branching ratios $BR$ are taken from the IAEA Live Chart of Nuclides: \url{https://www-nds.iaea.org/relnsd/vcharthtml/VChartHTML.html}} and the \ce{^{231}Pa} peak at \SI{5.159}{\MeV}: whereas in the previous work this range was completely unattributed, we can now identify the alpha decay of \ce{^{227}Ac} to \ce{^{223}Fr} ($BR=\SI{1.380}{\percent}$) as the most likely cause for a peak at \SI{5.042}{\MeV}, see \cref{fig:decomposition-high,fig:noSE-newLims-high-stacked-plot}.
With our current study, we find the most significant improvement in the reconstruction of the $\ce{^{234}U}$ peak at \SI{4.858}{MeV}~\footnote{The literature data for energy levels, half-lives $T_\mathrm{1/2}$, and branching ratios $BR$ are taken from the IAEA Live Chart of Nuclides: \url{https://www-nds.iaea.org/relnsd/vcharthtml/VChartHTML.html}} (\cref{fig:hypothesis-plot-comparison-high}).
Furthermore, we significantly improve the agreement between simulation and data in the energy range between the \ce{^{227}Th} peak at \SI{6.147}{\MeV} and the \ce{^{220}Rn} peak at \SI{6.405}{\MeV} (\cref{fig:decomposition-high}): here the LNM can attribute the tail to the right of \ce{^{227}Th} to \emph{individual alpha decays}\footnote{\ce{^{212}Bi} undergoes an alpha decay to \ce{^{208}Tl} in \SI{35.94}{\%} of the cases, otherwise a beta decay to \ce{^{212}Po}.} of \ce{^{212}Bi}, increasing the activity nearly two-fold, see \cref{tab:activityPerNuclide}, compared to the value assigned in the GFM based on SE assumptions. As the daughter nuclide \ce{^{212}Po} only has a half-life of \SI{294}{\ns}, a substantial amount of \ce{^{212}Bi} decays should be removed from the reference data as $\ce{^{212}Bi} \rightarrow \ce{^{212}Po}$ pile-up events, hence the obtained $A_\mathrm{212Bi}$ value can be only a lower limit. Since this would further heighten the excess over the SE-value, its plausibility is currently under investigation.

Similarly, the LNM can attribute the few events to the right of the \ce{^{211}Bi} peak at \SI{6.750}{\MeV} to \emph{individual decays} of \ce{^{219}Rn}, whereas the GFM can only calculate the \emph{total} \ce{^{219}Rn} activity, i.e.\ including pile-up events $\ce{^{219}Rn} \rightarrow \ce{^{215}Po}$ ($T_\mathrm{1/2, 215Po}=\SI{1.781}{\ms}$), from SE assumptions. Consequently, the value for $A_\mathrm{219Rn}$ found by LNM is well below the SE-values assumed in the GFM. We note that in \emph{total}, the \ce{^{219}Rn} activity may be in SE; however, the majority of its decays which are part of $\ce{^{219}Rn} \rightarrow \ce{^{215}Po}$ pile-ups is not accessible to the LNM as they are missing in the data.
%For \ce{^{214}Bi} and \ce{^{219}Rn}, the LNM finds activities well below the SE-values assumed in GFM (\cref{tab:activityPerNuclide}).
%This is to be expected given that those decays are removed from the reference data as $\ce{^{214}Bi} \rightarrow \ce{^{214}Po}$ ($T_\mathrm{1/2, 214Po}=\SI{163}{\us}$) and $\ce{^{219}Rn} \rightarrow \ce{^{215}Po}$ ($T_\mathrm{1/2, 215Po}=\SI{1.781}{\ms}$) pile-up events.

As the used background model considers only bulk contaminations, neither the GFM nor the LNM can reproduce the peak at \SI{5.304}{\MeV} that is caused by \ce{^{210}Po} contamination on the surface of the target crystal (see \cite{strauss2015beta} for details). Currently, we extend our background model to include surface contaminations, also to study the possibility that the tails around the \ce{^{238}U} peak (\SI{4.270}{\MeV}) and between the \ce{^{234}U} (\SI{4.857}{\MeV}) and \ce{^{231}Pa} (\SI{5.159}{\MeV}) peaks (see \cref{fig:noSE-newLims-high-stacked-plot}) may be caused by partial energy collection close to the surface.

\subsection{Radiogenic Beta and Gamma Background}\label{subsec:radiogenic-beta-gamma-bck}
Besides the dominant alpha lines, the background in the high energy range also has a contribution of beta/gamma decays of \ce{^{208}Tl}: due to the slow detector response of $\mathcal{O}(\si{\ms})$, beta decays of IR \ce{^{208}Tl} and subsequent gamma transitions to the ground state of \ce{^{208}Pb} cannot be resolved but are reconstructed as one event with a combined energy deposition up to \SI{\sim 4.9}{\MeV} (cf.\ \cref{fig:noSE-newLims-high-stacked-plot}). Whereas the GFM had to reconstruct the high energy range solely from alpha decaying contaminants, the LNM can also consider the featureless, high-energy tail of the \ce{^{208}Tl} template.

Up to an end-point of \SI{3.269}{\MeV}, the beta spectrum of individual \ce{^{214}Bi} decays are a prominent contribution (cf.\ \cref{fig:noSE-newLims-medium-stacked-plot}). Again, the GFM values are higher than the LNM ones (\cref{tab:activityPerNuclide}), as the former includes the pile-up events $\ce{^{214}Bi} \rightarrow \ce{^{214}Po}$ ($T_\mathrm{1/2, 214Po}=\SI{163}{\us}$).

Beta/gamma-decaying nuclides are the dominant source for background below the prominent photopeak of \ce{^{208}Tl} at \SI{2.615}{\MeV}. As shown in \cref{fig:noSE-newLims-roi-stacked-plot,fig:noSE-newLims-low-stacked-plot,fig:noSE-newLims-medium-stacked-plot}, each beta/gamma template has an individual endpoint and hence a different slope. This information is not usable with the GFM but gives the LNM a sensitivity to rather smooth, peak-less templates.
%This impacts clearly the IR background: also with LNM it stays the most prominent background category (\cref{tab:decomposition}) but now only below \SI{\sim 60}{\keV} (\cref{fig:decomposition-roi,fig:decomposition-low,fig:decomposition-medium}). This allows for a better fitting of the reference data between \SI{\sim 1800}{keV} and \SI{\sim 2000}{keV} (cf.\ \cref{fig:decomposition-medium}) by decreasing the activity of the IR background compared to GFM\@. Due to the improved background model (see \cref{appendixDiffToOld}), which was enabled by the LNM, the reproduction of flat background parts could be further improved in the \SIrange{1}{8}{keV} range, the \SIrange{25}{40}{keV} range, and the \SIrange{100}{300}{keV} range (cf.\ \cref{fig:decomposition-roi,fig:decomposition-low}). 
Together with the improved background model (see appendix~\cref{appendixDiffToOld}), which was enabled by the LNM, this allows for better reproduction of flat background parts in the \SIrange{1}{8}{keV}, the \SIrange{25}{40}{keV}, and the \SIrange{100}{300}{keV} (cf.\ \cref{fig:decomposition-roi,fig:decomposition-low}).

Below \SI{140}{\keV}, the background is dominated by IR components (cf.\ \cref{fig:decomposition-low}); in the ROI it reaches a coverage of \SI{48.4}{\percent}, see \cref{tab:decomposition}, with \ce{^{234}Th} being the most prominent contaminant. The NER background, caused by the detector holder made of NOSV copper, stays the least important background category within the ROI but increased to \SI{1.5}{\percent}.
This is mainly driven by higher activities for \ce{^{210}Pb}: with a $Q$-value of \SI{63.5}{\keV}, it contributes to the ROI and low energy range (\cref{fig:noSE-newLims-roi-stacked-plot,fig:hypothesis-plot-comparison-low}).
%This is mainly driven by higher activities for \ce{^{40}K} and \ce{^{210}Pb}: the former contributes significantly to the flat background below its photopeak at \SI{1.460}{\MeV} (see \cref{fig:noSE-newLims-medium-stacked-plot}) to the extent that the combined spectrum of NER background is dominated by the \ce{^{40}K} spectrum (\cref{fig:decomposition-medium}). With a $Q$-value of \SI{63.5}{\keV}, the latter contributes to the ROI and low energy range (\cref{fig:noSE-newLims-roi-stacked-plot,fig:hypothesis-plot-comparison-low}).
Preliminary results of ongoing screening measurements indicate a \ce{^{210}Pb} activity of $\mathcal{O}(\si{\milli\becquerel\per\kg})$, which is much higher than the fitted value of $A_\mathrm{210Pb,NER}=\SI{61(3)}{\micro\becquerel\per\kg}$. We found that the simplified geometry we applied in our current background model most likely causes a strong overestimation of the geometric efficiency $\varepsilon_\mathrm{g}$ (see \cref{eq:template}) for NER \ce{^{210}Pb}: an unrealistically low activity seems adequate to explain the observed background rate. Hence, the value stated in \cref{tab:activityPerNuclide} should be considered a lower limit. To mitigate this issue, a simulation with a more detailed geometry is currently in preparation and will be used in a future extended background model. 
We want to point out that because of the more precise screening information used in combination with the LNM (see \cref{tab:reference data}), nearly all other contaminants in the NER category, except for five nuclides (see \cref{tab:activityPerNuclide}), have activities that are either below $<\SI{0.01}{\micro\becquerel\per\kg}$ or compatible with zero within $2\sigma$. This leads to a decreased importance of NER background components in the medium energy range (\cref{fig:decomposition-medium}). 

Due to the stricter limits on NER contaminants, which we used for the LNM, the remaining gamma peaks in the medium energy range (\cref{fig:decomposition-medium,,fig:noSE-newLims-medium-stacked-plot}), which can be neither attributed to IR nor to NER, are now exclusively attributed to AER, resulting in an increased overall contribution of AER and making it the second most prominent background category in the ROI with an increased coverage of \SI{23.5}{\percent} (\cref{tab:decomposition}).
As stated in \cref{background}, we assume copper to be the most plausible source material for most of the AER background.

\subsection{Cosmogenic Background}
The IC background is present only in the low energy range and the contribution of it with its three constituents \ce{^3H}, \ce{^{179}Ta}, and \ce{^{181}W}, increased from \SI{17.6}{\percent} to \SI{20.5}{\percent} in the ROI (\cref{tab:decomposition}). Within $2\sigma$ around the respective mean values, the individual activities stay the same between the GFM and the LNM with one exception (\cref{tab:activityPerNuclide}). This is to be expected, as the sharp peaks associated with these contaminants make them equally suitable for both the GFM and the LNM\@. For the electron capture (EC) decay of \ce{^{179}Ta} the LNM, contrary to the GFM, is able to disentangle the three overlapping peaks due to EC from the M\textsubscript{1}, M\textsubscript{2}, and M\textsubscript{3} shells.
%Similar to the previously discussed case of \ce{^{40}K}, also here the GFM is not able to differentiate between templates with overlapping peaks.

The exception is \ce{^3H}, whose completely peak-less, pure beta spectrum with an endpoint at \SI{18.591}{keV} cannot be processed at all by the GFM due to its definition, i.e.\ relying on peaks. Hence, GFM-based background studies (\cite{turkolu2018development, CRESST:2019oqe}, appendix~\cref{appendixSEUpdate}) estimated the \ce{^3H} activity $A_\mathrm{3H}$ from ACTIVIA calculations to \SI{24.2(0.9)}{\micro\becquerel\per\kilo\gram}. In contrast, LNM-based studies, like this work, can naturally consider such peak-less spectra. To compensate for missing background at low energies, the LNM-based fit increased $A_\mathrm{3H}$ by almost a factor of three to \SI{64(4)}{\micro\becquerel\per\kilo\gram}, making up more than \SI{10}{\percent} of the flat background in the \SIrange{1}{20}{keV} range. Clearly, $A_\mathrm{3H}$ relies on the assumptions of the applied background model, especially on unknown contaminants at low energy, and hence may change in any future update of our model. Given that CRESST searches for light DM and thus a precise description of our background at low energies is crucial, further detailed careful studies of the \ce{^3H} contamination and the accuracy of $A_\mathrm{3H}$ are of high interest.

From extended ACTIVIA calculations~\cite{Cresst:2021arx}, we know already that we can expect further cosmogenically activated radioisotopes in our crystals, which will be introduced in a future extended background model.

\subsection{Correlations between Similar Contaminants}
\label{sec:correlations}

The sensitivity of the LNM to continuous parts of the spectral templates can allow it to disentangle background components caused by the same contaminant but originating from different volumes. Of course, the ability to do so depends on how much the continuous parts differ in shape and slope. 

\Cref{fig:40K-comparison} compares the template shapes from different geometric sources on the example of \ce{^{40}K}. Hence, we allowed independent \ce{^{40}K} contaminations in the NER, AER, and IR background categories. In contrast, the GFM left no choice but to attribute all \ce{^{40}K} contributions to one of the three background categories for which we arbitrarily chose AER. 

\begin{figure}[ht]
	\centering
	\includegraphics[width=0.6\linewidth]{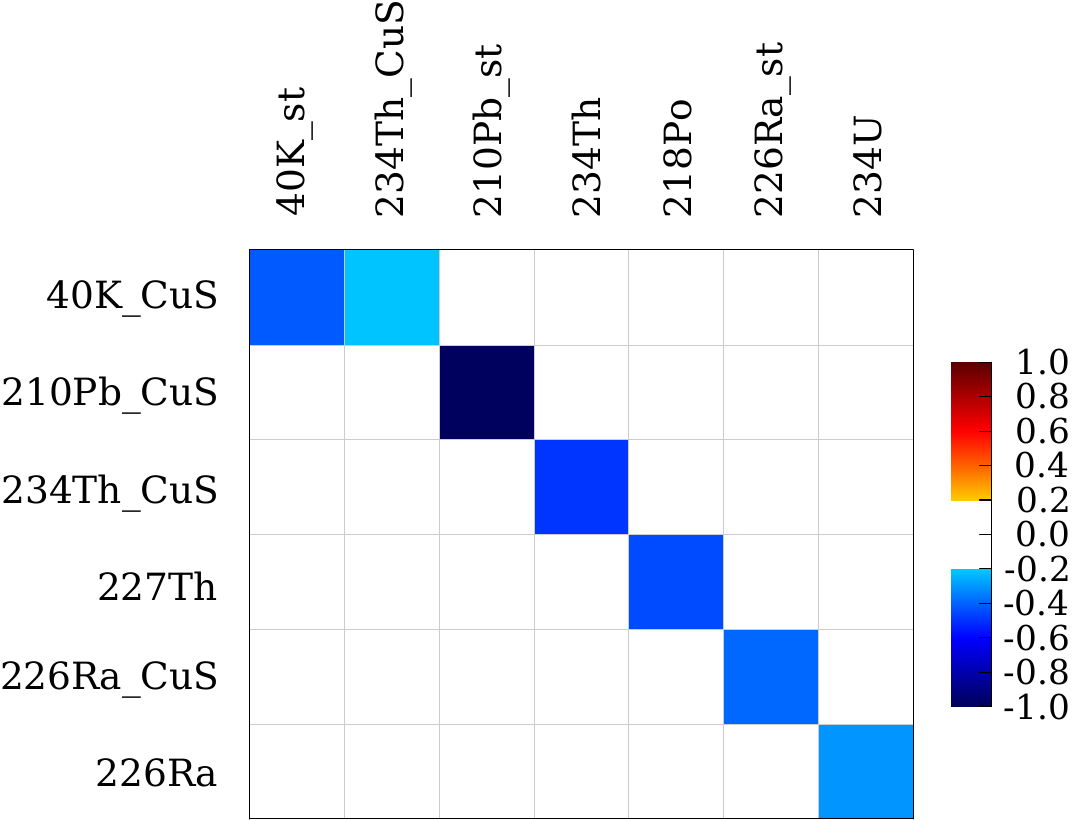}
	\caption{Correlations between parameters $\theta_k$ (see \cref{eq:expected-events}) of the likelihood normalisation method. Template names ending in \enquote{st} refer to NER components, while names ending in \enquote{CuS} refer to AER components. Only the part of the correlation matrix with absolute values larger than \num{0.2} is shown.}
	\label{fig:LNM-correlations}
\end{figure}

However, also the discrimination power of the LNM is limited: even though the LNM makes a firm attribution of \ce{^{40}K} solely to AER and none to NER and IR, we note the strong anti-correlation between the AER and the NER component, see \cref{fig:LNM-correlations}.
This means that the continuous spectral parts of the NER and AER components are too similar to allow for an unambiguous discrimination, see the inset in \cref{fig:40K-comparison} that shows the slight differences at low energies between the NER and AER templates.
On the other hand, a contribution from the IR component can be readily excluded since no significant anti-correlation can be observed; here, the spectrum is significantly different from the NER and AER templates, see \cref{fig:40K-comparison}.
A strong anti-correlation generally indicates uncertainty of the fit of which template to use since the corresponding parameters tend to complement each other.
Small changes in the starting conditions (like noise) might lead to significant parameter changes.

Further strong anti-correlations are observed for \ce{^{210}Pb}, \ce{^{226}Ra}, and \ce{^{234}Th} when the same nuclide is placed in different volumes, and for \ce{^{40}K}/\ce{^{234}Th}, \ce{^{218}Po}/\ce{^{227}Th}, and \ce{^{226}Ra}/\ce{^{234}U} when these nuclides are placed in the same volume, see \cref{fig:LNM-correlations}. The anti-correlation of \ce{^{40}K}/\ce{^{234}Th} comes from the similar continuous spectral shapes paired with the prominent copper fluorescence peak around \SI{8.0}{keV} in both templates (see \cref{fig:noSE-newLims-roi-stacked-plot}). Overlapping peaks in the alpha region cause the anti-correlation between \ce{^{226}Ra}/\ce{^{234}U} and \ce{^{218}Po}/\ce{^{227}Th} (see \cref{fig:noSE-newLims-high-stacked-plot}).

\subsection{Overall Spectral Agreement}
The spectral agreement of the LNM-based background fit to the reference data provides a better description of the background than the GFM\@. 
The explainable percentages (see \cref{eq:explainable-percentage}) are improved in each energy range separately; for the combined range, it is improved by \SI{7.9}{\percent}. As shown in the hypothesis plot \cref{fig:hypothesis-plot-comparison-high}, we now better reproduce the \ce{^{223}Ra} peak at \SI{5.978}{\MeV} in the high energy range. In the ROI (\cref{fig:hypothesis-plot-comparison-roi}), the peaks around \SI{2.6}{keV} and \SI{11}{keV} caused by EC decays of \ce{^{179}Ta} from the M and L shells, respectively, match better. Also, the tension between the model and data around \SI{25}{\keV} and below \SI{5}{\keV} is mended, the latter due to the increased \ce{^{3}H} contribution. In the low and medium energy ranges, the LNM enables a better reproduction of the continuous parts of the observed background, resulting in less tension in the hypothesis plots between \SIrange{120}{220}{\keV} and \SIrange{500}{800}{\keV}, see \cref{fig:hypothesis-plot-comparison-low,fig:hypothesis-plot-comparison-medium}.

There are still tensions at \SI{\sim 13}{\keV}, \SI{\sim 37.5}{\keV}, and \SI{\sim 41.8}{\keV}, where our model cannot reproduce peaks. In an attempt to explain the peak at \SI{\sim 13}{\keV}, we added \ce{^{228}Ra}, which features a gamma line at \SI{13.52}{\keV}, to the AER component of our model, allowing essentially an unconstrained contribution of \ce{^{228}Ra}. However, the LNM-based fit set $A_\mathrm{228Ra,AER}<\SI{0.01}{\micro\becquerel\per\kg}$ (\cref{tab:activityPerNuclide}) and therefore excludes it as a source for the observed peak. 
These missing peaks and the remaining difference of the flat background between \SIrange{140}{240}{keV} (\cref{fig:hypothesis-plot-comparison-low,fig:noSE-newLims-low-stacked-plot}) suggest that there might be additional background sources that were not considered so far, e.g.\ contaminants in the scintillating and reflecting polymeric foil that is placed in the detector modules. Work is going on to extend our future background model addressing this issue. 

%Regarding the spectral agreement, we note that the diverging width of the \ce{^{208}Tl} peak (at \SI{2.615}{\MeV} in \cref{fig:noSE-newLims-medium-stacked-plot}) is not the result of a wrongly implemented energy resolution in our detector response model. We currently investigate whether this difference is caused by the approximated source geometry for AER contaminants.

\subsection{Secular Equilibrium Assumption}

\begin{figure*}[ht!]
	\centering
      \begin{subfigure}[b]{\textwidth}
    	\centering
    	\includegraphics[width=1.0\linewidth]{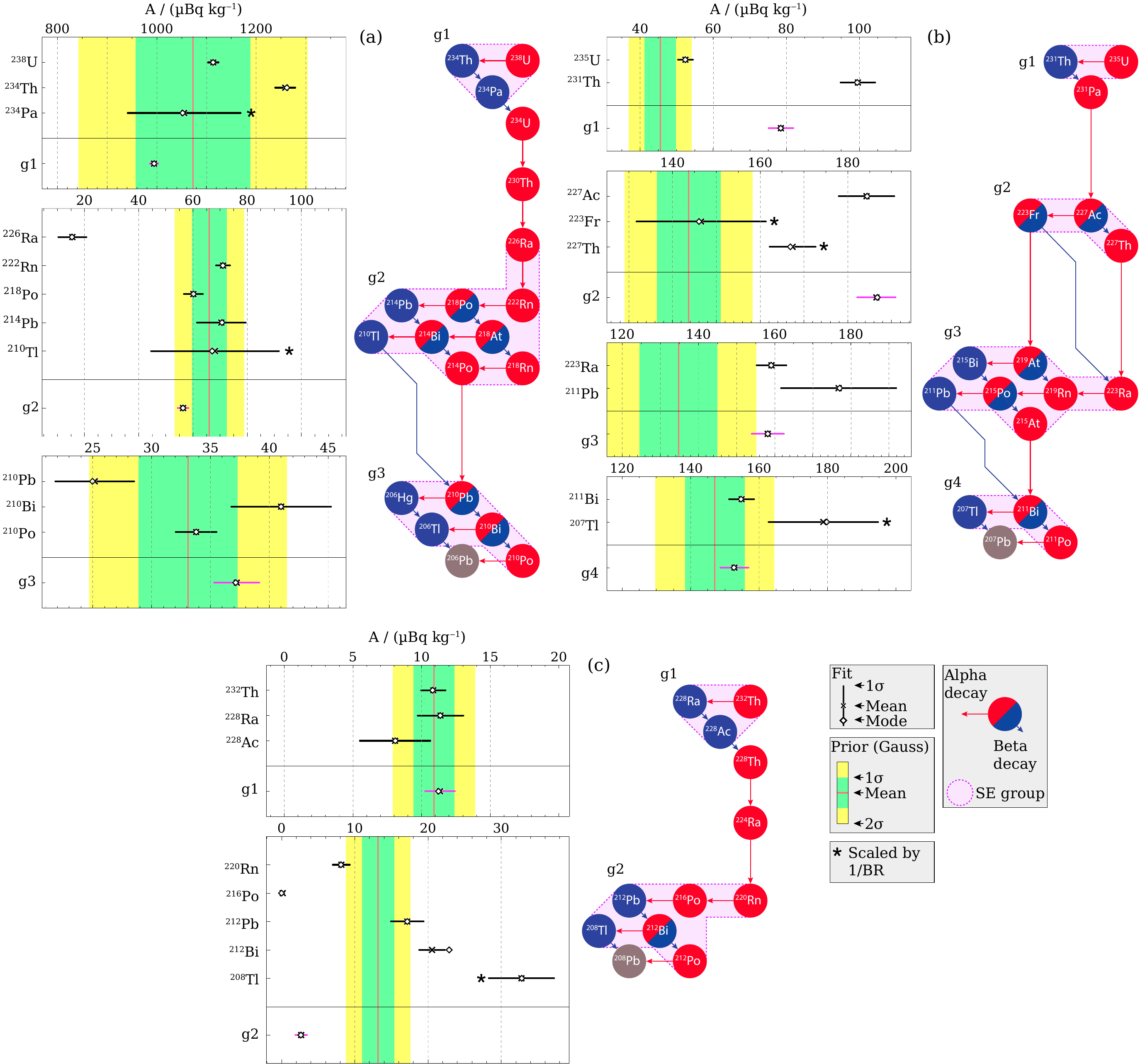}
    	\phantomcaption\label{fig:se-assessment:238U}
        \phantomcaption\label{fig:se-assessment:235U}
        \phantomcaption\label{fig:se-assessment:232Th}
    \end{subfigure}
	\caption{SE assessment of IR components from (\protect\subref{fig:se-assessment:238U}) the $^{238}\mathrm{U}$ decay chain, (\protect\subref{fig:se-assessment:235U}) the $^{235}\mathrm{U}$ decay chain, and (\protect\subref{fig:se-assessment:232Th}) the $^{232}\mathrm{Th}$ decay chain. The full decay chains are shown at the \textit{right sides}, \textit{pink} areas enclose SE groups $\mathrm{g}_i$.
    At the \textit{left sides}, the SE groups' activities scaled to the respective parent nuclide contributions are shown as \textit{purple bars} (SE fit), while \textit{blue bars} show the activities of the groups' constituents (uncorrelated fit). Activities marked with an asterisk are scaled by the inverse branching ratio to make them comparable to their parent nuclides' activities. Similarly, the SE groups' activities are scaled to the groups' parent activities. Figure adapted from \cite{CRESST:2022rez}, see there for details about the SE fit.
    }
	\label{fig:se-assessment}
\end{figure*}

Leaving the 84 components of the LNM unconstrained by SE assumptions raises the question about the physical plausibility of the obtained results. Due to the time elapsed between modifying the radiochemical composition of the crystals during production and the start of data taking, one can assume that at least some sub-groups of the natural decay chains retain secular equilibrium.
However, the purpose of this work is to demonstrate the technical capabilities of the LNM and its potential to cross-check SE assumptions with respect to the GFM since assuming no SE allows the LNM to compensate for unidentified or unused background components. In addition, as we showed in~\cite{CRESST:2022rez}, even without imposing partial SE a priori, LNM recovers most of the partial SE a posterior.

We concluded that only three of the nine SE groups in the \cawo{} crystal show deviation from SE and that the reason for these deviations is an effect of the likelihood normalisation since the half-lives of these nuclides are too short to have an impact on the data. \Cref{fig:se-assessment} reports the IR activities of all SE groups in comparison to the activities of a correlated fit, i.e.\ when enforcing strict SE assumptions. The three aforementioned groups violating SE are: the second group of the $^{238}\mathrm{U}$ chain (g\textsubscript{2} in \cref{fig:se-assessment:238U}), the first group of the $^{235}\mathrm{U}$ chain (g\textsubscript{1} in \cref{fig:se-assessment:235U}), and the second group of the $^{232}\mathrm{Th}$ chain (g\textsubscript{2} in \cref{fig:se-assessment:232Th}).

In the first case, $^{238}\mathrm{U}$/g\textsubscript{2}, the LNM returns a broken equilibirum between $^{226}\mathrm{Ra}$ and $^{222}\mathrm{Rn}$, see \cref{fig:se-assessment:238U}, which is implausible given the short half-life of $T_{1/2}=\SI{3.82}{\day}$ for $^{222}\mathrm{Rn}$. This is a technical artefact caused by the strong anti-correlation of $^{226}\mathrm{Ra}$ with $^{234}\mathrm{U}$, see \cref{fig:LNM-correlations} and \cref{sec:correlations}. 

In case of $^{235}\mathrm{U}$/g\textsubscript{1}, the broad and peak-less \ce{^{231}Th} template is used by the fit to compensate for the gaps between \SI{100}{keV} and \SI{300}{keV} (see \cref{fig:noSE-newLims-low-stacked-plot}), whereas its parent nuclide \ce{^{235}U} is restricted from the alpha region.

In the last case, $^{232}\mathrm{Th}$/g\textsubscript{2}, the spectral templates from alpha-decaying nuclides overlap with other alpha peaks (\ce{^{220}Rn}/\ce{^{211}Bi}, \ce{^{212}Bi}/\ce{^{227}Th}, \ce{^{216}Po}/\ce{^{219}Rn}, see also \cref{fig:noSE-newLims-high-stacked-plot}) and the ones from beta-decaying nuclides are peak-less and broad (\ce{^{212}Bi}, \ce{^{208}Tl}, \ce{^{212}Pb}, see also \cref{fig:noSE-newLims-low-stacked-plot,,fig:noSE-newLims-medium-stacked-plot}).

In the first group of the $^{238}\mathrm{U}$ chain, the activities of all nuclides are within $2\sigma$ of the prior and are therefore considered to be in agreement with SE.\@ The second group of the $^{235}\mathrm{U}$ chain shows that lower activities are attributed to the daughters $^{228}\mathrm{Fr}$ and $^{227}\mathrm{Th}$ in comparison to the mother nuclide $^{227}\mathrm{Ac}$. However, it can still be considered to be consistent with SE expectations, as the group will only be in full secular equilibrium after $\sim\SI{30}{yr}$ (see \cite[fig.\ 4.9]{burkhart2022enhancing}).
Based on the results from~\cite{CRESST:2022rez}, we currently study the impact of different strictness levels of SE constraints on the results of the LNM approach. An obvious candidate for an imposed SE condition would be the $^{226}\mathrm{Ra} \rightarrow ^{222}\mathrm{Rn}$ decay.

\begin{table}[h]
	\centering
	\caption{Comparison of coverages (see \cref{eq:coverage}) between the normalisation using the Gaussian fit method (\textit{GFM})~(appendix~\cref{appendixSEUpdate}) and this work's likelihood normalisation method (\textit{LNM}) in the different energy ranges (see \cref{background}). The improvement (\textit{Impr.}) measures the change towards \SI{100}{\percent}: $|\zeta_{j,\mathrm{GFM}} - \SI{100}{\percent}|-|\zeta_{j,\mathrm{LNM}} - \SI{100}{\percent}|$}
	\begin{tabular}{lccccc}
		\toprule
		    & ROI        & Low        & Medium      & High        & Sum   \\
		\midrule[.08em]
		GFM      & \SI{74.6}{\percent} & \SI{74.6}{\percent} & \SI{88.7}{\percent} & \SI{97.8}{\percent} & \SI{84.0}{\percent} \\
		LNM & \SI{93.9}{\percent} & \SI{94.6}{\percent} & \SI{110.4}{\percent} & \SI{100.2}{\percent} & \SI{99.6}{\percent} \\
		\midrule
		Impr.    & \SI{19.3}{\percent} & \SI{20.0}{\percent} & \SI{0.9}{\percent}   & \SI{2.0}{\percent} & \SI{15.6}{\percent} \\
		\bottomrule
	\end{tabular}
	\label{tab:comparison-coverage-cenk-noSE-newLims}
\end{table}

\begin{table}[h]
	\centering
	\caption{Comparison of explainable percentages (see \cref{eq:explainable-percentage}) between the normalisation using the Gaussian fit method (\textit{GFM})~(appendix~\cref{appendixSEUpdate}) and this work's likelihood normalisation method (\textit{LNM}) in the different energy ranges (see \cref{background}). The improvement (\emph{Impr.}) measures the change towards \SI{100}{\percent}: $|\mathrm{EP}_{j,\mathrm{GFM}} - \SI{100}{\percent}|-|\mathrm{EP}_{j,\mathrm{LNM}} - \SI{100}{\percent}|$}
	\begin{tabular}{lccccc}
		\toprule
		    & ROI        & Low        & Medium     & High       & Sum   \\
		\midrule[.08em]
		GFM      & \SI{64.5}{\percent} & \SI{77.2}{\percent} & \SI{88.7}{\percent} & \SI{34.4}{\percent} & \SI{68.2}{\percent} \\
		LNM & \SI{82.7}{\percent} & \SI{88.1}{\percent} & \SI{90.9}{\percent} & \SI{41.6}{\percent} & \SI{76.1}{\percent} \\
		\midrule
		Impr.    & \SI{18.6}{\percent} & \SI{10.9}{\percent} & \SI{2.2}{\percent}  & \SI{7.2}{\percent} & \SI{7.9}{\percent} \\
		\bottomrule
	\end{tabular}
	\label{tab:comparison-explainable-percentage-cenk-noSE-newLims}
\end{table}

\begin{figure*}
	\centering
	\includegraphics[width=1.0\textwidth]{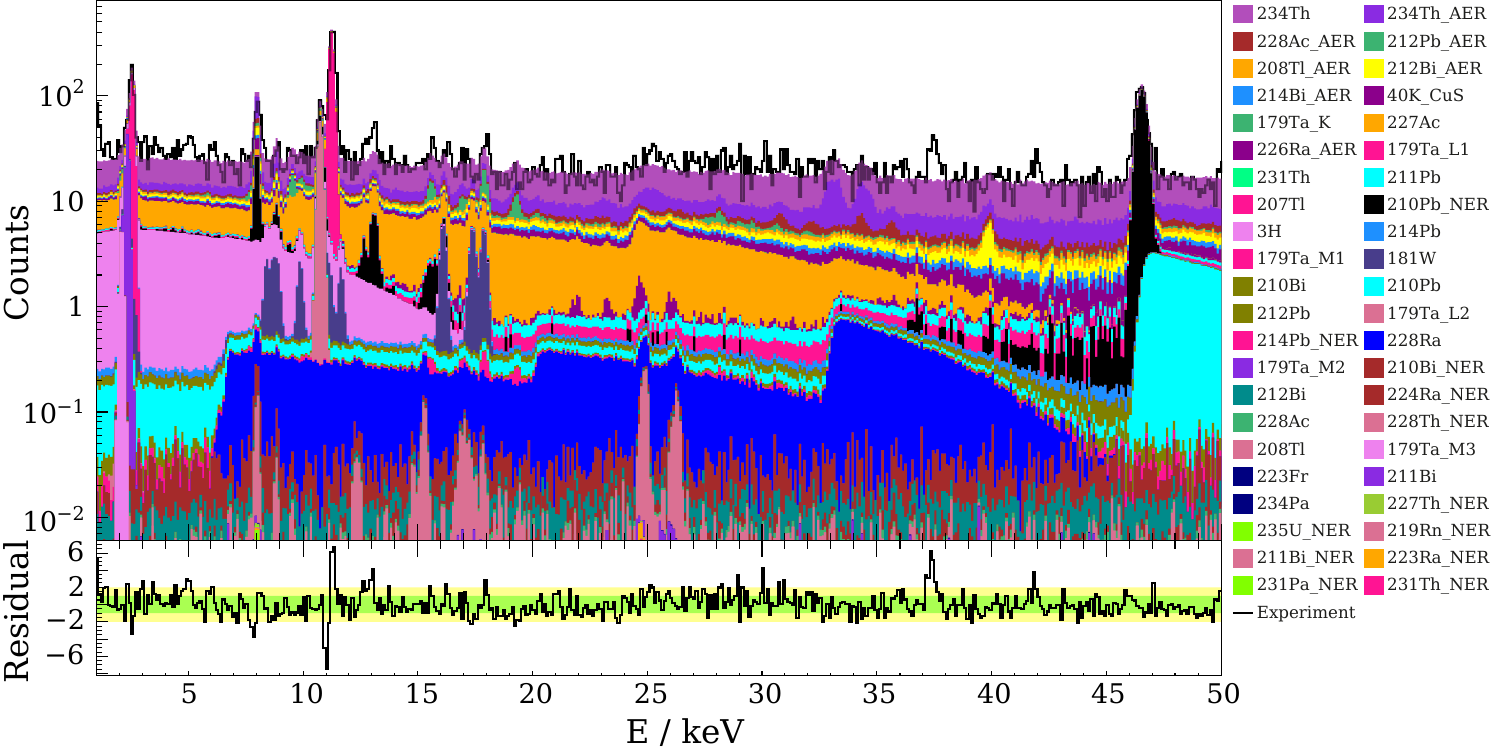}
	\caption{Stacked plot of all templates in the ROI compared to the reference data (\textit{black line}) with a bin width of \SI{100}{\eV}. The legend is ordered by the nuclides' coverages from highest (\textit{upper left}) to lowest (\textit{lower right}); an interactive version of this plot is provided as Online Resource 1.}
	\label{fig:noSE-newLims-roi-stacked-plot}
\end{figure*}

%\subsection{Componet Decomposition}
%IR
%changes >~3sig
%226Ra alpha droped
%214Bi alpha/beta droped 
%219Rn alpha drop
%216Po alpha drop
%208Tl beta drop
%231Th beta gain
%227Ac alpha/beta gain
%212Bi alpha/beta gain (above SE)
%-> activity loss by betas: 279 uBq/kg
%   activity gain by betas: 108 uBq/kg
%=> seems strange that coverage increase if net activity droped

%NER
%Except 40K, 214Pb, 210Pb, 210Bi, 228Th, 224Ra all nuclides drop <0.01 or are compatible with 0 at 2sig
%Gain by 40K (+190 uBq/kg) can't compensate loss by beta-nuclide 234Th, 234Pa, 214Pb, 210Pb, 210Bi that drops from 5*123uBq/kg=619uBq/kg to 81uBq/kg
%=> seems strange that coverage increase so much when net activity droped

%AER
%234Th overcompensate loss in NER, IR stable
%228Ac net loss wrt to NER and IR

%IC
%stable except rise in 3H and now disentangle 179Ta(M2)

\begin{table}[h]
	\centering
	\caption{Comparing the coverages of the individual background components in the energy range of interest between the normalisation using the Gaussian fit method (\textit{GFM})~(appendix~\cref{appendixSEUpdate}, \cref{tab:comparison}) as baseline and
this work’s likelihood normalisation method (\textit{LNM}). Values are rounded to one decimal place}
	\begin{tabular}{lcc}
		\toprule
		Component & GFM & LNM \\
		\midrule[.08em]
		Internal radiogenic  & 40.7\%         & 48.4\%              \\
		Internal cosmogenic  & 17.6\%         & 20.5\%              \\
		Near external radiogenic & 0.2\%          & 1.5\%               \\
		Additional external radiogenic& 15.9\%         & 23.5\%              \\
		\midrule
		Sum                  & 74.6\%         & 93.9\%              \\
		\bottomrule
	\end{tabular}
	\label{tab:decomposition}
\end{table}

In conclusion, while having completely free-floating parameters might not be the physically most accurate model (it is unlikely that all SE groups are broken), it can be interesting to see the full potential of the simulated data.
This can help to assess if an unexplained energy region under partial SE assumptions is truly absent in the current simulations or if a particular template might be able to fill the gap but is held back by SE constraints. 
If the latter case is true, two likely explanations are conceivable: SE is truly broken, or an additional source from a different CRESST setup component is missing in the simulations. 
In practice, this can only be determined by further simulations of a more detailed setup geometry, by considering more contaminants, and/or through dedicated screening measurements like the one carried out in this work (see \cref{copper}).

\begin{figure*}
	\centering
	\includegraphics[width=1.0\textwidth]{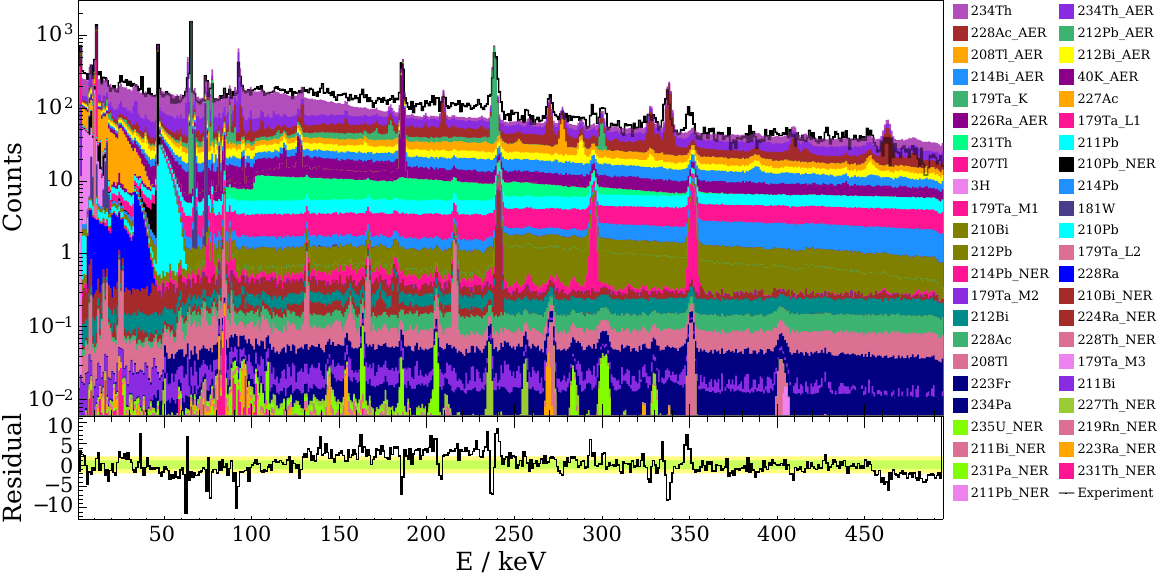}
	\caption{Stacked plot of all templates in the low energy range compared to the reference data (\textit{black line}) with a bin width of \SI{1}{\keV}. The legend is ordered by the nuclides' coverages from highest (\textit{upper left}) to lowest (\textit{lower right}); an interactive version of this plot is provided as Online Resource 2.}
	\label{fig:noSE-newLims-low-stacked-plot}
\end{figure*}

\begin{figure*}
	\centering
	\includegraphics[width=1.0\textwidth]{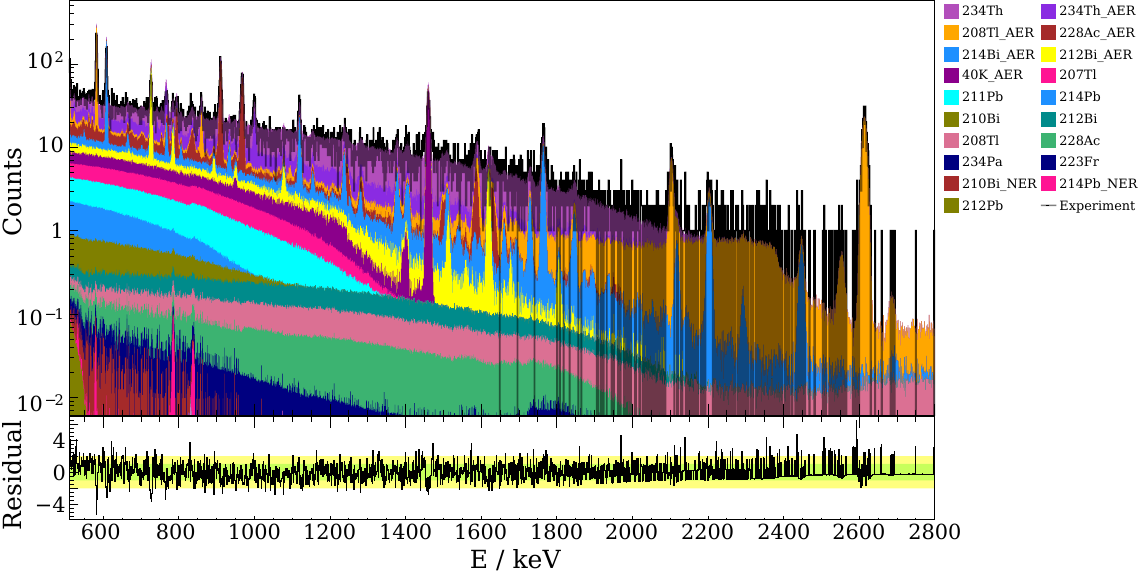}
	\caption{Stacked plot of all templates in the medium energy range compared to the reference data (\textit{black line}) with a bin width of \SI{1}{\keV}. The legend is ordered by the nuclides' coverages from highest (\textit{upper left}) to lowest (\textit{lower right}); an interactive version of this plot is provided as Online Resource 3.}
	\label{fig:noSE-newLims-medium-stacked-plot}
\end{figure*}

\begin{figure*}
	\centering
	\includegraphics[width=1.0\textwidth]{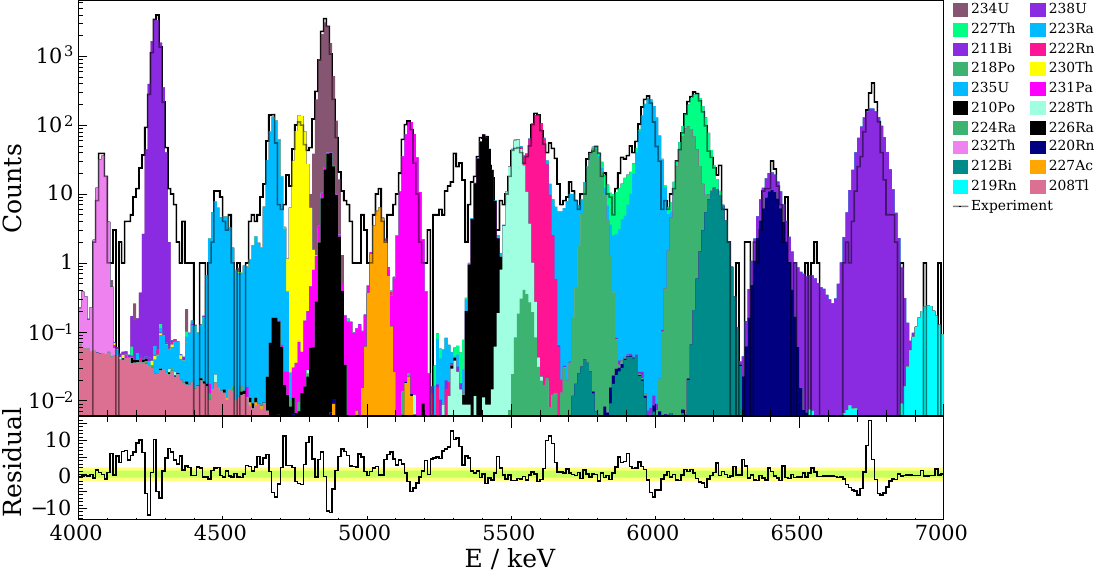}
	\caption{Stacked plot of all templates in the high energy range with a bin width of \SI{10}{\keV}. The legend is ordered by the nuclides' coverages; an interactive version of this plot is provided as Online Resource 4. Possible sources of the events in the blank regions are currently being investigated.}
	\label{fig:noSE-newLims-high-stacked-plot}
\end{figure*}

\begin{figure*}
    \centering
    \begin{subfigure}[b]{0.475\textwidth}
        \centering
    	\includegraphics[width=1.0\linewidth]{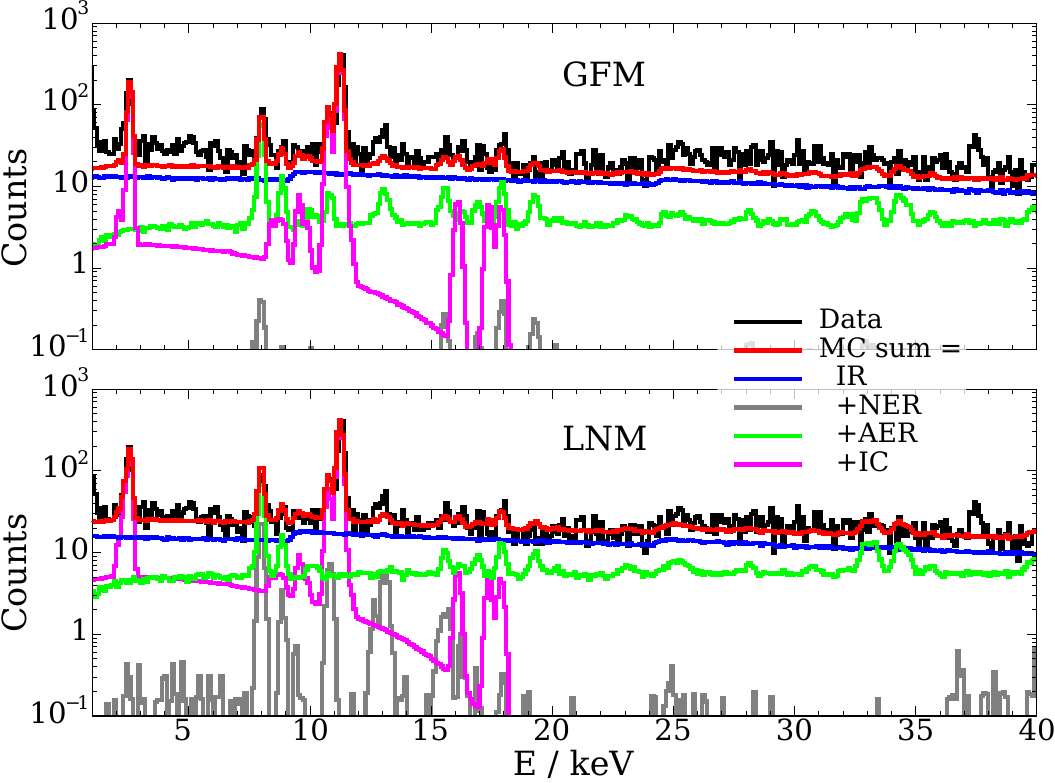}
    	\caption{ROI}
    	\label{fig:decomposition-roi}
    \end{subfigure}
    \hfill
    \begin{subfigure}[b]{0.475\textwidth}
    	\centering
    	\includegraphics[width=1.0\linewidth]{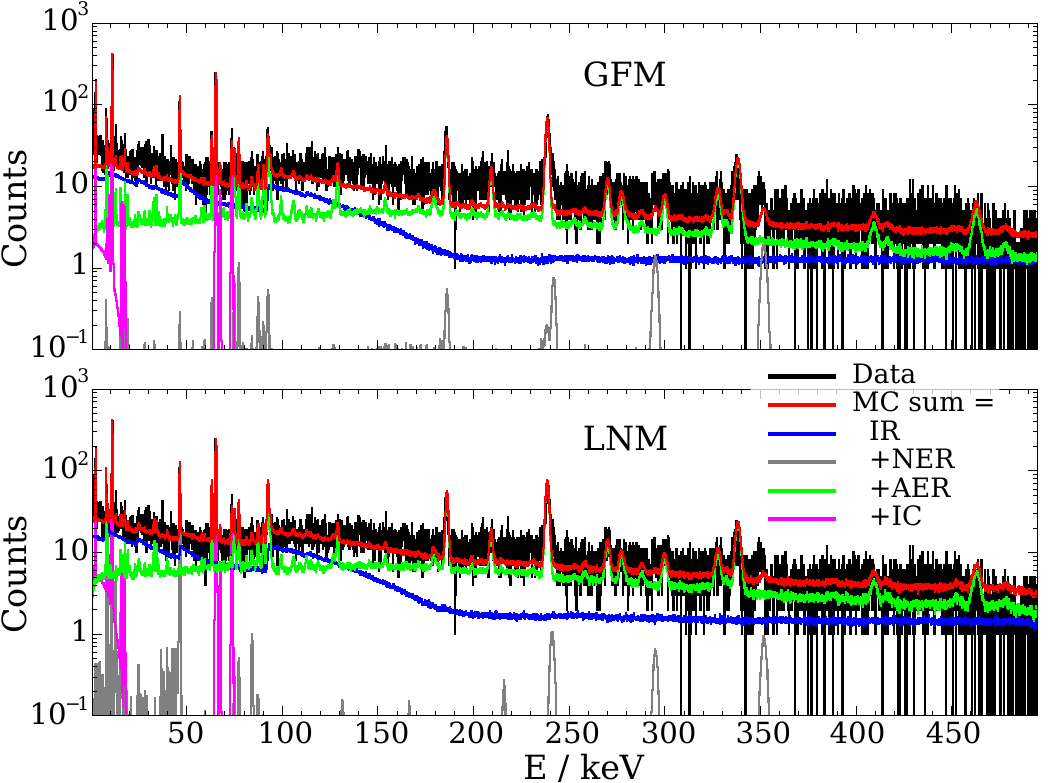}
    	\caption{Low energy range}
    	\label{fig:decomposition-low}
    \end{subfigure}
    \vskip\baselineskip
    \begin{subfigure}[b]{0.475\textwidth}
    	\centering
    	\includegraphics[width=1.0\linewidth]{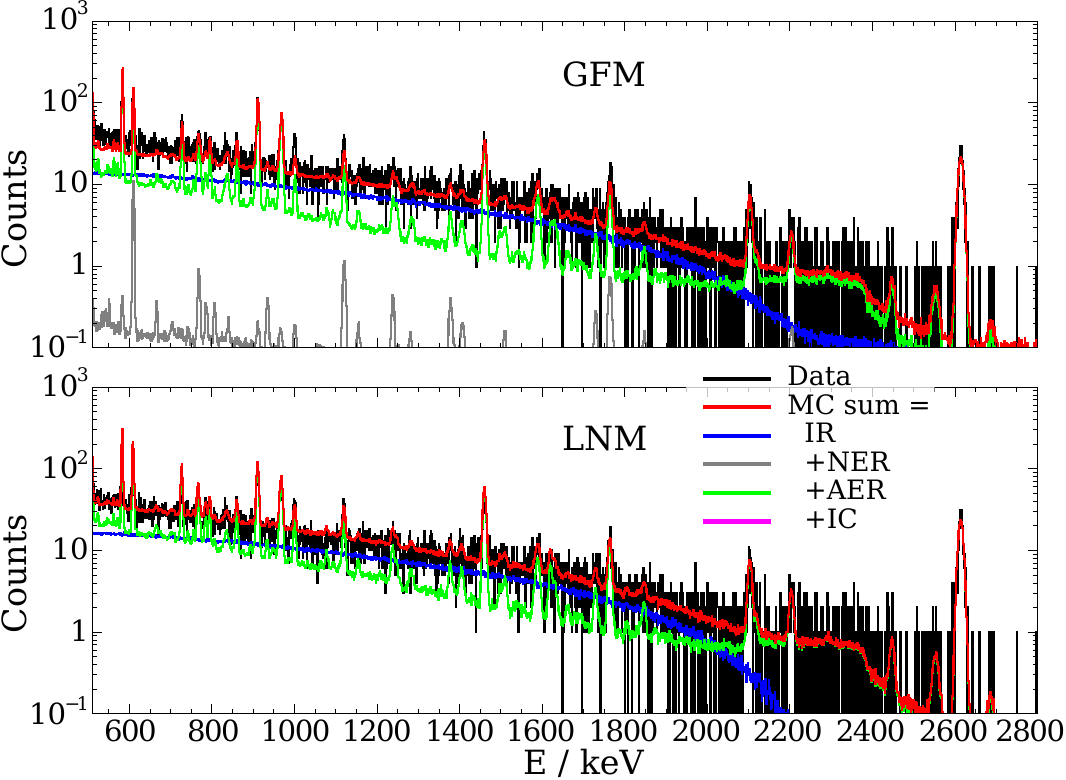}
    	\caption{Medium energy range}
    	\label{fig:decomposition-medium}
    \end{subfigure}
    \hfill
    \begin{subfigure}[b]{0.475\textwidth}
    	\centering
    	\includegraphics[width=1.0\linewidth]{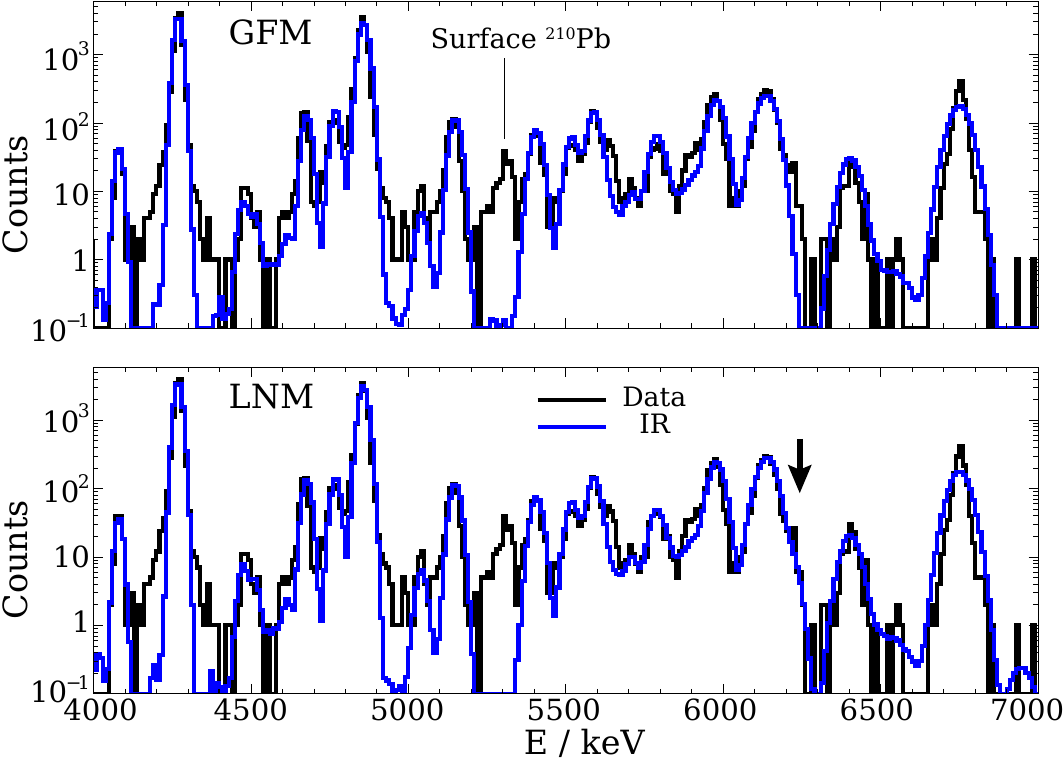}
    	\caption{High energy range}
    	\label{fig:decomposition-high}
    \end{subfigure}

    \caption{Decomposition of the fitted background into \emph{internal radiogenic} (IR, \emph{blue} line), \emph{near external radiogenic} (NER, \emph{gray} line), \emph{additional external radiogenic} (AER, \emph{green} line), and \emph{internal cosmogenic} (IC, \emph{pink} line) in comparison to the experimental reference data (\emph{black} histogram) for (\protect\subref{fig:decomposition-roi}) the ROI with \SI{100}{eV} binning, (\protect\subref{fig:decomposition-low}) the low energy range with \SI{100}{eV} binning, (\protect\subref{fig:decomposition-medium}) the medium energy range with \SI{1}{keV} binning, and (\protect\subref{fig:decomposition-high}) the high energy range with \SI{10}{keV} binning. The performance of the updated \emph{Gaussian fit method}~(GFM, see appendix~\cref{appendixSEUpdate}) is shown in the respective \textit{top} panel and the newly developed \emph{likelihood normalisation method} (LNM) in the respective \textit{bottom} panel. Note that the spectral shapes of the fits in (\protect\subref{fig:decomposition-high}) only appear to be identical but features improved around the peak of \ce{^{212}Bi} which is indicated by a \textit{black arrow}; also the \ce{^{210}Po} surface contamination is not considered in the current background model.}
    
    \label{fig:decompositions}
\end{figure*}

\section{Summary and outlook} \label{summary_outlook}
In this work, we introduced the LNM as a technically more powerful method to estimate the electromagnetic background of the CRESST experiment from experimental measurements on the example of the TUM40 detector module.
Compared to the previously used GFM, the newly adopted LNM, based on a Bayesian likelihood fit, features three major benefits: 
(i) it is independent of the explicit assumption of partial SE a priori, (ii) it can handle peak-less background components (e.g.\ \ce{^3H}), and (iii) it can disentangle partially overlapping peaks (e.g.\ EC peaks of \ce{^{179}Ta}). Overall, it shows an enhanced capability to reproduce the observed continuous background. To improve the priors for the likelihood fit, we performed dedicated measurements of the radioactive contamination levels in the copper used for the TUM40 detector holder in the frame of an ongoing screening campaign. The results are the most sensitive limits obtained for NOSV copper from this batch, which shows the apparent overestimation of this component in the previous model.

Within the applied background model, consisting of 84 individual background components without enforced partial SE, the LNM 
%Using a Bayesian likelihood fit with 84 components, we
could potentially cover %\SI{90.6 +- 0.9}{\percent}
\SI{93.9}{\percent} of the observed electromagnetic background in the TUM40 detector module in the ROI\@. This is a major improvement compared to the %\SI{73.9+-7.9}{\percent} 
\SI{74.6}{\percent} obtained with the previously used method.
%; it is an increase of \SI{19.3}{\percent}.
The explainable percentage -- an intuitive GOF measure introduced in this work -- grew from \SI{64.5}{\percent} to \SI{82.7}{\percent} at a \num{0.01} nominal significance. 
%This is an increase by \SI{18.6}{\percent}.

This is partially driven by a significantly higher attribution of background to a potential \ce{^3H} contamination in the \cawo{} crystal: \SI{64(4)}{\micro\becquerel\per\kg} compared to \SI{24.2(0.9)}{\micro\becquerel\per\kg} obtained with the GFM method.
Further studies are needed to assess how robust these results and their physic interpretations are against changes in the underlying background model, e.g.\ by adding more background components or imposing various degrees of SE~\cite{burkhart2022enhancing}.
%In the \cawo{} crystal of TUM40, we found a significantly higher \ce{^3H} activity of \SI{68.8 +- 4.1}{\micro\becquerel\per\kg} compared to \SI{24.0 +- 0.9}{\micro\becquerel\per\kg} obtained with the GFM method. Further studies are needed to assess if this increase is of physical origins or an artefact of the likelihood fit.

At the moment, we are actively working on improvements to our future background model. Extensive material assays and screening campaigns are planned to provide more precise priors for the likelihood normalisation. Furthermore, we are extending our background model: we are investigating additional contaminants %and surface background
as potential origins of so far unidentified peaks at \SI{13}{\keV}, \SI{37.5}{\keV}, and \SI{41.8}{\keV} (see \cref{fig:noSE-newLims-low-stacked-plot}) and unattributed alpha background in the \SI{4}{\MeV} to \SI{7}{\MeV} range (see \cref{fig:noSE-newLims-high-stacked-plot}). These additional contaminants consist of surface contaminations, further cosmogenics~\cite{Cresst:2021arx}, and contributions from experiment components not considered thus far, like the scintillating foils inside the detector modules. The approximated source geometry for the AER background is currently replaced with a detailed geometry of the cryostat. From this, we expect a more precise determination of the \ce{^{210}Pb} activity in external copper parts of the setup and an impact on the spectral template.
%of e.g. external \ce{^{208}Tl}.
We are also switching to a more recent Geant4 version. As Geant4 is under continuous development, this will give us access to the most recent improvements that may affect the simulation of our templates.
%and finalising a model for our neutron background.
As another step, we plan to apply the likelihood normalisation method to further detector modules and different target materials, e.g.\ \ce{Al_2O_3} and \ce{LiAlO_2}.

%\backmatter

%\bmhead{Supplementary information}
%This article has four accompanying supplementary files.

\acknowledgments
The computational results presented were partially obtained using the CLIP cluster and the Max Planck Computing and Data Facility (MPCDF).
This work has been funded by the Deutsche Forschungs\-gemeinschaft (DFG, German Research Foundation) under Germany's Excellence Strategy – EXC 2094 – 390783311 and through the Sonderforschungsbereich (Collaborative Research Center) SFB1258 ‘Neutrinos and Dark Matter in Astro- and Particle Physics’, by the BMBF 05A20WO1 and 05A20VTA and by the Austrian science fund (FWF): I5420-N, W1252-N27. FW was supported through the Austrian research promotion agency (FFG), project ML4CPD. SG was supported through the FWF project STRONG-DM (FG1). JB and HK were funded through the FWF project P 34778-N ELOISE. The Bratislava group acknowledges a partial support provided by the Slovak Research and Development Agency (projects APVV-15-0576 and APVV-21-0377). 

%\bmhead{Conflict of interest}
%The authors have no relevant financial or non-financial interests to disclose.

%\item Ethics approval 
%\item Consent for publication
%\bmhead{Data Availability}
%The data sets generated and analysed for the presented study are available from the corresponding authors on reasonable request.

%\bmhead{Code availability}
%The codes used are currently under preparation for publication. Till then, they are available from the corresponding authors on reasonable request.

%\bmhead{Author contributions}
%Concept of the likelihood normalisation method and design of this study was developed by HK and VM. HK and JB devised the used likelihood formulation. JB implemented the likelihood normalisation method as code, conducted the fit, and analysed the results. VM organized the material screening campaign; FF, ML, SN, and VM prepared and executed the screening measurements. The manuscript was written by JB, HK, and VM. All authors read and approved the final manuscript.
\appendix
%\section{Some title}
%Please always give a title also for appendices.
%\section{Updated Results using the Gaussian Fitting Method}\label{appendixSEUpdate}
%\begin{appendices}

\section{Updated Results using the Gaussian Fitting Method}\label{appendixSEUpdate}
%From the TUM40 erratum
The original implementation of the GFM in~\cite{CRESST:2019oqe} contains a technical shortcoming in the analysis code. 
It caused an underestimation of \ce{^{234}Th} in the IR background component and overall an overestimation of the NER background components, especially \ce{^{210}Pb}, \ce{^{218}At}, \ce{^{226}Ra}, \ce{^{228}Ra}, \ce{^{228}Th}, and \ce{^{234}Pa}. Both effects nearly cancel each other out, leading to an insignificant impact on the total sum in \cref{tab:comparison} with respect to the uncertainties.

To provide a reliable base to which the new likelihood normalisation presented in this work can be compared, we provide here updated results of the GFM normalisation. Updated background spectra \cite[figs.11, 12]{CRESST:2019oqe} are given in \cref{fig:decompositions}, updated contamination activities \cite[table 4]{CRESST:2019oqe} and relative contributions \cite[table 5]{CRESST:2019oqe} per background component, and activities per nuclide \cite[tables 7 to 11]{CRESST:2019oqe} are given in \cref{tab:sum_activity,tab:comparison,tab:activityPerNuclide}, respectively.

Besides the update, we used the same improved energy resolution for the spectral template as for the LNM, see appendix~\cref{appendixDiffToOld}. Beyond these, no changes to the physics and methodology outlined in \cite{CRESST:2019oqe} were applied. Especially, the NER components are still normalised to the contamination values reported by CUORE \cite{alduino2016cuore} and \textit{not} to those reported in \cref{copper} for our own copper.

%Based on this example: https://tex.stackexchange.com/questions/305395/table-alignment-around-pm-sign-how-to-have-good-alignment-without-pm-where-a
\begin{table*}
\begin{threeparttable}
\centering
\caption{Activities $A$ for individual background components over the full energy range and the resulting background rate $R$ in the ROI according to the Gaussian fit method. Statistical uncertainties at \SI{68}{\percent} C.L., values rounded to two significant digits on the smallest uncertainty}
\label{tab:sum_activity}
\begin{tabular*}{\textwidth}{@{\extracolsep{\fill}}lll@{\hskip 3pt}c@{\hskip 3pt} S[table-format=1.4(4), separate-uncertainty]}
\toprule
\multicolumn{3}{l}{Component}   & {$A \pm \delta_\mathrm{stat} \pm \delta_\mathrm{sys}$ / \si{\micro\becquerel\per\kg}} & {$R \pm (\delta_\mathrm{stat} + \delta_\mathrm{sys})$ / \si{\per\kg\per\keV\per\day}} \\ \midrule
\multicolumn{3}{l}{Internal} & & \\
& \multicolumn{2}{l}{Radiogenic} & & \\
& & $^{238}\mathrm{U}$ decay chain     &  3781  $\pm$ 26 $\pm$ 110 & 0.634 +- 0.011\\
& & $^{235}\mathrm{U}$ decay chain     &  1256.7 $\pm$ 8.2 $\pm$ 63.5 & 0.2326 +- 0.0079\\
& & $^{232}\mathrm{Th}$ decay chain    &   155.2 $\pm$ 1.1 $\pm$ 20.7 & 0.0184 +- 0.0026\\
& \multicolumn{2}{l}{Cosmogenics}      &   448.27 $\pm$ 0.63 $\pm$  9.23 & 0.383 +- 0.010  \\\midrule
\multicolumn{3}{l}{External radiogenic} & & \\
& \multicolumn{2}{l}{Near\tnote{a}}             &   774.71 $\pm$ 0.73 & {$(5.186 \pm 0.069)\cdot 10^{-3}$}\\ 
& \multicolumn{2}{l}{Additional}       &  3432.0 $\pm$ 4.9 $\pm$ 369.3 & 0.346 +- 0.035\\ \midrule
\multicolumn{3}{l}{Total sum} & 9848 $\pm$ 38 $\pm$ 573 & 1.619 +- 0.066\\
\bottomrule
\end{tabular*}
\begin{tablenotes}
     \item[a] In this case, no systematic uncertainty can be given, see \cite{CRESST:2019oqe} for details.
\end{tablenotes}
\end{threeparttable}
\end{table*}

\begin{table}
\centering
\caption{Relative contribution for individual background components in the ROI according to the Gaussian fit method. Linearly combined statistical (at \SI{68}{\percent} C.L.) and systematic uncertainties; values rounded to two significant digits on the uncertainty}
\label{tab:comparison}
\begin{tabular*}{\columnwidth}{@{\extracolsep{\fill}} ll@{\hskip 3pt} S[table-format=2.1(2), separate-uncertainty] S[table-format=2.4(4), separate-uncertainty]}
\toprule
\multicolumn{2}{l}{\multirow{2}{*}{Component}}     & \multicolumn{2}{c}{Relative contribution (\%)}\\
\multicolumn{2}{l}{}                               & {Original work \cite{CRESST:2019oqe}} & {This work} \\
\midrule
\multicolumn{2}{l}{Internal} & & \\
& Radiogenic            & 26.6 +- 5.0                                & 40.7   +-   1.9\\
& Cosmogenic            & 17.8 +- 3.8                                & 17.63   +-   0.57\\
\multicolumn{2}{l}{External radiogenic} & & \\
& Near                  &  6.3 +- 2.1                                &  0.2388 +-   0.0028\\
& Additional            & 17.5 +- 4.9                                & 15.9   +-   3.2\\ \midrule
\multicolumn{2}{l}{Total sum}                      & 68.2 +- 15.8                               & 74.6   +-   5.6\\
\bottomrule
\end{tabular*}
\end{table}

\section{Improvements of the Background Model}\label{appendixDiffToOld}
The used background model was originally developed in \cite{CRESST:2019oqe} and further improved for this work. The changes are:

Because the original background model was the first ever developed for CRESST, the IR background component still contained radionuclides, which afterwards could be omitted as it turned out that their contribution to the background is negligible because their branching ratio is insignificant and only minimal leakage to lower energies occurs (see \cref{tab:activityPerNuclide} for details). Furthermore, in our previous work, we normalised the background model solely with the GFM, which lacked the capability of the LNM to discriminate between templates that share the same peaks but deviate in their continuous parts. Hence, in the previous work, all activity of \ce{^{40}K} had to be assigned to the AER background component and could not be allowed to be shared between AER, NER, and IR\@.

Because the GFM cannot handle featureless, pure beta spectra, we could not directly normalise the \ce{^3H} contribution to the IC background component. Instead, it was related to the activity of \ce{^{181}W} via ACTIVIA~\cite{BACK2008286} calculations. Furthermore, as the GFM has only limited capability to process overlapping peaks, we could not individually normalise the spectral templates of the \ce{^{179}Ta} M-shells but had to treat it as one combined template. As the used Geant4 version wrongly applied the same branching ratio to all shells, this was a systematic shortcoming. In total, the previous background model contained only six instead of nine IC components.

Similar to the IR background, the selected 28 radionuclides for the NER background category were different from the original model. As the complete observed background activity of \ce{^{40}K} was originally assigned to AER, the total number of components for NER was only 28. Originally, the activities of these radionuclides were normalised to screening results from the CUORE experiment \cite{alduino2016cuore}, which used the same batch of copper as CRESST, under the assumption of complete SE through each of the three radioactive decay chains. This was necessary because contaminations by the same radionuclide but in different volumes, here the \cawo{} target (IR background) and the detector module holder (NER background), can cause similar spectral templates which differ not in their peaks but in their continuous part. Hence, with the GFM it was not possible to normalise the NER components to the reference data in the original work, whereas in this work with the LNM, it is possible.

Finally, the AER background originally consists of only nine observed nuclides: \ce{^{40}K}, \ce{^{208}Tl}, \ce{^{210}Pb}, \ce{^{212}Pb}, \ce{^{212}Bi}, \ce{^{214}Bi}, \ce{^{226}Ra}, \ce{^{228}Ac}, and \ce{^{234}Th}. Only after publication, \ce{^{228}Ra} was identified as a possible source for a yet unidentified peak at \SI{13}{\keV}. As the precise origin of the AER components is still unknown, we continue to use the "Cu shell approximation", which we developed in~\cite{CRESST:2019oqe}.

Overall, our original background model contained 88 individual background components in four categories with 30 free parameters, compared to 84 individual components in the current work with 84 free parameters.

\section{Screening Methodology}\label{appendixscreening}
\subsection{Copper Sample Preparation} \label{subsec:screening}
%\subsubsection{Sample preparation} \label{subsubsec:sample}
The samples, in the form of plates, were cut from several big copper sheets kept in a storage cellar of the Weihenste\-phan brewery %\footnote{\url{https://www.weihenstephaner.de/en/our-brewery/brewing-process/}} 
(\SI{15}{\m} deep) close to Munich to avoid cosmo\-genic activation of the material.

We prepared a sample with a total mass of \SI{95.71}{\kg} that consists of two types of plates: 18 plates without and 16 with a hole in the centre to fill the cavity of the HPGe detector as much as possible. Their size is $\SI{245}{mm}\times\SI{245}{mm}\times\SI{8}{mm}$ and the hole, if present, is \O$\SI{100}{mm}\times\SI{8}{mm}$. In each plate, four threaded smaller holes were drilled to facilitate placing them inside the sample chamber of the HPGe spectrometer. Only one plate without a hole was used for alpha spectrometry. The sample for the ICP-MS measurement was a small piece cut from one of the plates with a total mass of \SI{34.85}{\gram} and a size of $\SI{6}{mm}\times\SI{10}{mm}\times\SI{108}{mm}$.

\begin{figure}[ht]\centering
\includegraphics[width=0.8\linewidth]{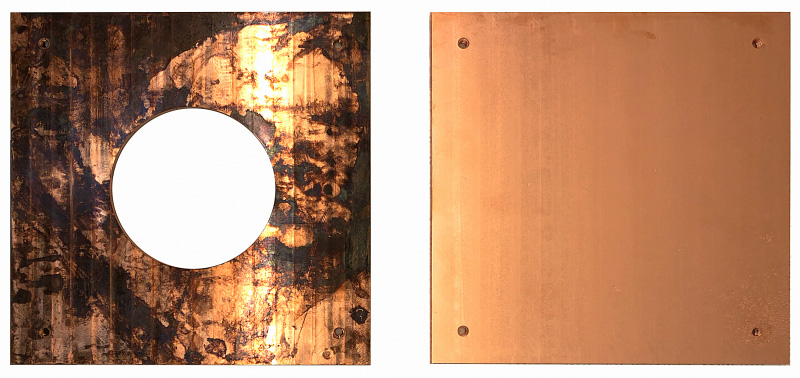}
\caption{Photos of NOSV copper samples before (\textit{left}) and after (\textit{right}) chemical surface cleaning.}
\label{fig:copper_clean_dirty}
\end{figure}

The surface oxidation caused by long-time exposure to air was removed by mechanical grinding of the copper surface. Around \SI{0.5}{mm} of the top layer was polished off. To further minimise the radioactive contamination of the surfaces introduced during the sample cutting and handling, the following cleaning procedure was applied:
\begin{itemize}
    \item [$-$] cleaning with acetone;
    \item [$-$] ultrasonic bath in a mixture of alkaline cleaning solution and water;% (ratio is not critical);
    \item [$-$] rinsing with high purity DI-water (resistivity value is \SI{18.2}{\mega\ohm\per\cm} at \SI{25}{\celsius});
    \item [$-$] etching in a water solution of nitric acid (\SI{70}{\percent});
    \item [$-$] passivation in a water solution of hydrogen peroxide (ratio 1:1) and citric acid (\SI{5}{\percent} of mass).
\end{itemize}
All used chemicals were of ultra-pure quality. In \cref{fig:copper_clean_dirty}, the difference before and after the cleaning of the plates is shown.

\subsection{Gamma Ray Spectrometry with HPGe Detectors} \label{subsubsec:hpge}
Gamma-ray spectrometry performed with germanium detectors allows us to directly quantify the activity of many radionuclides contributing to the background of an experiment. The measurement was performed in the underground low background laboratory STELLA (SubTErranean Low Level Assay) at LNGS~\cite{Arpesella,Laubenstein:2017yjj}. The p-type coaxial germanium detector~\cite{NEDER2000191,HEUSSER2006495,Rugel:2005svs} used for the measurement is shielded with $20\, \mathrm{cm}$ of lead (with low level of $\mathrm{^{210}Pb}$) and \SI{5}{\cm} of NOSV copper. The sample chamber with a volume of \SI{0.0138}{\m\cubed} is constantly flushed with boil-off nitrogen from a nearby liquid nitrogen storage tank. 
%The total background rate of the detector is $44\,/( \mathrm{kg \cdot day})$ integral between \num{40} and  \SI{2700}{keV}. 
The total background rate of the detector between \num{40} and \SI{2700}{keV} is \SI{44}{\per\kg\per\day}. 
The energy resolution of the detector is \SI{2.63}{keV} (FWHM) for the \SI{1332.52}{\keV} gamma-line of $\mathrm{^{60}Co}$. The data for the copper sample were accumulated in the energy range of \SIrange{50}{3000}{keV} during \SI{2540}{\hour}. 

\subsection{Inductively Coupled Plasma Mass Spectrometry} \label{subsubsec:icp-ms}
To remove the surface contamination due to handling and cutting, the copper sample was treated with several partial etchings by \SI{20}{\milli\liter} of $4\, \mathrm{M}$ (Molar) solution of ultrapure nitric acid, obtained using a sub-boiling distillation system\footnote{DuoPur, Milestone, Italy, \url{https://www.milestonesrl.com/products/clean-chemistry/subpur-and-duopur}}. About \SI{3}{\gram} of copper were so\-lubilised every etching. All solutions were stored sepa\-rately.

After about $70\mathrm{\%}$ of the copper mass was dissolved, the solution of the last etching was collected in perfluoroalkoxy (PFA) vials previously cleaned and checked. This solution was divided into two aliquots: the first was used for the measurement of the contamination in the bulk of the sample; the second was spiked with \SI[per-mode=symbol]{25}{\pico\gram\per\gram} of $\mathrm{Th}$ and $\mathrm{U}$, using a reference standard solution\footnote{AccuStandard, New Haven, USA, \url{https://www.accustandard.com}}, to evaluate the recovery yield of the sample treatment procedure.\\
Subsequently, a matrix separation was carried out in co\-lumns filled with TRU resin\footnote{Triskem, France, \url{https://www.triskem-international.com/catalog/products/resins-and-accessories/tru-resin/bl,product,417,0}} to pre-con\-centrate $\mathrm{Th}$ and $\mathrm{U}$. Conditioning of the columns was performed by alternate rinsing steps using $4\, \mathrm{M}$ nitric acid and $0.1\, \mathrm{M}$ ammonium oxalate. %($\mathrm{C_{2}H_{8}N_{2}O_{4}}$).%
Then the $4\, \mathrm{M}$ acid solutions containing the metal were loaded in the columns where the analytes were eluted with \SI{10}{\milli\liter} of $0.1\,\mathrm{M}$ ammonium oxalate.

The concentration of $\mathrm{Th}$ and $\mathrm{U}$ was measured using Sector Field-ICP-MS\footnote{Element II, Thermo Fisher Scientific, USA, \url{https://www.thermofisher.com/order/catalog/product/IQLAAEGAAMFABWMAFC}}. All the sample treatment procedures and measurements were performed inside an ISO 6 class clean room at LNGS to reduce the risk of possible environmental contamination. 
The concentration of $\mathrm{Th}$ and $\mathrm{U}$ measured in the copper sample was $<\SI[per-mode=symbol]{0.5e-12}{\gram\per\gram}$ for both isotopes.
Using conversion factors\footnote{\url{https://radiopurity.org}}, we receive the activities listed in \cref{tab:reference data}.

\section{Obtained Activities}\label{appendixFullResults}
\Cref{tab:activityPerNuclide} lists the activity values per background component as obtained with the updated GFM (see appendix~\cref{appendixSEUpdate}) and the LNM\@.
Some components, marked with \emph{a}, are strongly suppressed by branching ratios $\lesssim \SI{0.01}{\percent}$ in their respective branch. As the previous work \cite{CRESST:2019oqe} and the updated GFM proved their insignificance for the background model, they are omitted from the LNM.

Some components, marked with \emph{b}, undergo a pure alpha decay with a $Q$-value beyond the upper energy limit of \SI{7}{\MeV} of the used reference data (see \cref{sec:refdata}). In the previous work, the GFM assigned activities by assuming partial SE with a component within the reference data. These components were not considered because we dismiss SE assumption in this work. We note that it is trivially possible to consider them with the LNM using the same SE assumption.
For the GFM, alpha decaying nuclides are excluded if they do not contribute to the background in the ROI\@.

For components marked with \emph{c}, a considerable pile-up with their fast-decaying daughter nuclide happens. As these pile-ups are excluded from the reference data, only the individual, not piled-up decays remain and can be fitted. Hence, the stated activities are only \emph{lower limits} on the total activity of these components. 

For all components, except those of the AER, the activity is given at the source of the background. For the AER components, the activity is given as measured by the detector. Since the origin of the AER background is not known precisely, the specific source activity cannot be calculated.

Because the GFM uses upper limits from CUORE \cite{alduino2016cuore} for the copper contamination level, only upper limits can be given for NER components. 

\renewcommand{\arraystretch}{1.4}
\onecolumn
\begin{center}

\begin{ThreePartTable}
\begin{longtable}{lllcc}
\caption{Results of the Gaussian fit method (GFM) and likelihood normalisation method (LNM) per background component in terms of contamination activities ($A$). The results of the GFM in this work (see appendix~\cref{appendixSEUpdate}) are updated with respect to the original results \cite{CRESST:2019oqe}, see the text for details. The total uncertainty is the linear sum of statistical uncertainties (due to the sample statistic at \SI{68}{\percent} C.L.) and systematic uncertainties (due to the normalisation values). Values are rounded to one significant digit on the uncertainty\label{tab:activityPerNuclide}}\\% "\\" are needed see https://tex.stackexchange.com/questions/148061/longtable-misplaced-noalign
%
% Header of the first page of the table
\toprule
 \multicolumn{3}{c}{\multirow{2}{*}{Background component}} & \multicolumn{2}{c}{($A \pm \delta_\mathrm{stat+sys}$) / \si{\micro\becquerel\per\kg}} \\ %\cmidrule{3-5}%Seems cmidrule is broken in overleaf, it gives error
 \multicolumn{3}{c}{} & GFM & LNM \\ \midrule
\endfirsthead
%
% Header of any subsequent pages
\multicolumn{5}{c}%
{\tablename\ \thetable\ - \textit{Continued from previous page}} \\
\toprule

 \multicolumn{3}{c}{\multirow{2}{*}{Component}} & \multicolumn{2}{c}{($A \pm \delta_\mathrm{stat+sys}$) / \si{\micro\becquerel\per\kg}} \\ %\cmidrule{3-5}%Seems cmidrule is broken in overleaf, it gives error
 \multicolumn{3}{c}{} & GFM & LNM \\ \midrule
\endhead
%
% Footer of all pages except the last
\bottomrule
\multicolumn{5}{r}{\textit{Continued on next page}} \\
\endfoot
%
% Footer of the last page
\bottomrule
\endlastfoot
%
% Table body
    \multicolumn{5}{l}{Internal Radiogenic (IR)} \\
     & & \ce{^{40}K} & -- & $<0.01$ \\
     & \multicolumn{4}{l}{\ce{^{238}U} chain} \\
     & & \ce{^{238}U}  & $1070 \pm 30$ & $ 1110 \pm 10 $ \\
     & & \ce{^{234}Th} & $1070 \pm 20$ & $ 1260 \pm 20 $ \\
     & & \ce{^{234}Pa} & $1.70 \pm 0.03$ & $ 1.7 \pm 0.2 $ \\
     & & \ce{^{234}U}  & $1090 \pm 40$ & $ 1150 \pm 10 $ \\
     & & \ce{^{230}Th} & $51 \pm 5$ & $ 52 \pm 2 $ \\
     & & \ce{^{226}Ra} & $66 \pm 6$ & $ 15 \pm 5 $ \\
     & & \ce{^{222}Rn} & $66 \pm 6$ & $ 71 \pm 3 $ \\
     & & \ce{^{218}Po} & $66 \pm 6$ & $ 60 \pm 3 $ \\
     & & \ce{^{214}Pb} & $66 \pm 5$ & $ 71 \pm 9 $ \\
     & & \ce{^{218}At} & $0.013 \pm 0.001$ & --\tnote{a} \\
     & & \ce{^{218}Rn} & --\tnote{a} & --\tnote{a}\\
     & & \ce{^{214}Bi} & $66 \pm 5$ & $<0.01$\tnote{c} \\
     & & \ce{^{210}Tl} & $0.014 \pm 0.001$ & $ 0.014 \pm 0.006 $ \\
     & & \ce{^{214}Po} & $66 \pm 5$ & --\tnote{b} \\
     & & \ce{^{210}Pb} & $33 \pm 3$ & $ 25 \pm 4 $ \\
     & & \ce{^{206}Hg} & $(63 \pm 5)\cdot 10^{-6}$ & --\tnote{a} \\
     & & \ce{^{210}Bi} & $33 \pm 3$ & $ 41 \pm 4 $ \\
     & & \ce{^{206}Tl} & $0.0044 \pm 0.0004$ & --\tnote{a} \\
     & & \ce{^{210}Po} & $33 \pm 3$ & $ 34 \pm 2 $ \\
     & \multicolumn{4}{l}{\ce{^{235}U} chain} \\
     & & \ce{^{235}U}  & $46 \pm 4$ & $ 52 \pm 2 $ \\
     & & \ce{^{231}Th} & $46 \pm 3$ & $ 99 \pm 4 $ \\
     & & \ce{^{231}Pa} & $45 \pm 4$ & $ 49 \pm 2 $ \\
     & & \ce{^{227}Ac} & $144 \pm 5$ & $ 184 \pm 6 $ \\
     & & \ce{^{223}Fr} & $1.98 \pm 0.07$ & $ 2.0 \pm 0.2 $ \\
     & & \ce{^{227}Th} & $140 \pm 10$ & $ 166 \pm 5 $ \\
     & & \ce{^{223}Ra} & $140 \pm 10$ & $ 159 \pm 4 $ \\
     & & \ce{^{219}At} & $(124  \pm 4) \cdot 10^{-6}$ & --\tnote{a} \\
     & & \ce{^{215}Bi} & $(113  \pm 4) \cdot 10^{-6}$ & --\tnote{a} \\
     & & \ce{^{219}Rn} & $135 \pm 6$ & $ 0.2 \pm 0.2 $\tnote{c} \\
     & & \ce{^{215}Po} & $135 \pm 6$ & --\tnote{b} \\
     & & \ce{^{211}Pb} & $135 \pm 6$ & $ 180 \pm 10 $ \\
     & & \ce{^{215}At} & $0.031 \pm 0.001$ & --\tnote{b} \\
     & & \ce{^{211}Bi} & $147 \pm 9$ & $ 155 \pm 4 $ \\
     & & \ce{^{207}Tl} & $147 \pm 6$ & $ 180 \pm 10 $ \\
     & & \ce{^{211}Po} & $0.41 \pm 0.02$ & --\tnote{b} \\
     & \multicolumn{4}{l}{\ce{^{232}Th} chain} \\
     & & \ce{^{232}Th} & $11 \pm 2$ & $ 10.8 \pm 0.9 $ \\
     & & \ce{^{228}Ra} & $11 \pm 2$ & $ 11 \pm 2 $ \\
     & & \ce{^{228}Ac} & $11 \pm 2$ & $ 8 \pm 2 $ \\
     & & \ce{^{228}Th} & $25 \pm 4$ & $ 30 \pm 2 $ \\
     & & \ce{^{224}Ra} & $31 \pm 4$ & $ 25 \pm 2 $ \\
     & & \ce{^{220}Rn} & $13 \pm 2$ & $ 8 \pm 1 $ \\
     & & \ce{^{216}Po} & $13 \pm 2$ & $<0.01$ \\
     & & \ce{^{212}Pb} & $13 \pm 2$ & $ 17 \pm 2 $ \\
     & & \ce{^{212}Bi} & $13 \pm 2$ & $ 23 \pm 5 $\tnote{c} \\
     & & \ce{^{208}Tl} & $4.7 \pm 0.7$ & $ 12 \pm 2 $ \\
     & & \ce{^{212}Po} & $8 \pm 1$ & --\tnote{b} \\
    \multicolumn{5}{l}{Near External Radiogenic (NER)} \\
     & & \ce{^{40}K} & -- & $<0.01$ \\
     & \multicolumn{4}{l}{\ce{^{238}U} chain} \\
     & & \ce{^{238}U}  & -- & --\tnote{a} \\
     & & \ce{^{234}Th} & $<123.88$ & $<0.01$ \\
     & & \ce{^{234}Pa} & $<0.1982$ & $ 0.1 \pm 0.1 $ \\
     & & \ce{^{234}U}  & -- & --\tnote{a} \\
     & & \ce{^{230}Th} & -- & --\tnote{a} \\
     & & \ce{^{226}Ra} & $<123.9$ & $<0.01$ \\
     & & \ce{^{222}Rn} & -- & --\tnote{a}\\
     & & \ce{^{218}Po} & -- & --\tnote{a}\\
     & & \ce{^{214}Pb} & $<123.9$ & $ 6.2 \pm 0.1 $\\
     & & \ce{^{218}At} & $<0.02478$ & $<0.01$ \\
     & & \ce{^{218}Rn} & --\tnote{a} & --\tnote{a}\\
     & & \ce{^{214}Bi} & $<123.88$ & $<0.01$ \\
     & & \ce{^{210}Tl} & $<0.02602$ & $<0.01$ \\
     & & \ce{^{214}Po} & -- & --\tnote{a} \\
     & & \ce{^{210}Pb} & $<123.88$ & $ 61 \pm 3 $ \\
     & & \ce{^{206}Hg} & $<\num{2.354e-6}$ & --\tnote{a} \\
     & & \ce{^{210}Bi} & $<123.88$ & $ 6.2 \pm 0.9 $ \\
     & & \ce{^{206}Tl} & $<\num{1.6588e-4}$ & --\tnote{a} \\
     & & \ce{^{210}Po} & -- & --\tnote{a} \\
     & \multicolumn{4}{l}{\ce{^{235}U} chain} \\
     & & \ce{^{235}U}  & -- & $ 0.04 \pm 0.04 $ \\
     & & \ce{^{231}Th} & $<0.8672$ & $ 0.04 \pm 0.02 $ \\
     & & \ce{^{231}Pa} & $<0.8672$ & $ 0.04 \pm 0.03 $ \\
     & & \ce{^{227}Ac} & -- & --\tnote{a} \\
     & & \ce{^{223}Fr} & $<0.01197$ & $<0.01$ \\
     & & \ce{^{227}Th} & $<0.855$ & $ 0.04 \pm 0.04 $ \\
     & & \ce{^{223}Ra} & $<0.867$ & $ 0.04 \pm 0.03 $ \\
     & & \ce{^{219}At} & -- & --\tnote{a} \\
     & & \ce{^{215}Bi} & $<\num{6.721e-7}$ & $<0.01$ \\
     & & \ce{^{219}Rn} & $<0.867$ & $ 0.04 \pm 0.04 $ \\
     & & \ce{^{215}Po} & -- & --\tnote{a} \\
     & & \ce{^{211}Pb} & $<0.867$ & $ 0.04 \pm 0.02 $ \\
     & & \ce{^{215}At} & -- & --\tnote{a} \\
     & & \ce{^{211}Bi} & $<0.8672$ & $ 0.04 \pm 0.04 $ \\
     & & \ce{^{207}Tl} & $<0.8648$ & $ 0.04 \pm 0.04 $ \\
     & & \ce{^{211}Po} & -- & $<0.01$ \\
     & \multicolumn{4}{l}{\ce{^{232}Th} chain} \\
     & & \ce{^{232}Th} & -- & $<0.01$ \\
     & & \ce{^{228}Ra} & $<3.812$ & --\tnote{a} \\
     & & \ce{^{228}Ac} & $<3.812$ & $<0.01$ \\
     & & \ce{^{228}Th} & $<3.812$ & $ 2.0 \pm 0.2 $ \\
     & & \ce{^{224}Ra} & $<3.812$ & $ 2.00 \pm 0.06 $ \\
     & & \ce{^{220}Rn} & -- & --\tnote{a} \\
     & & \ce{^{216}Po} & -- & --\tnote{a} \\
     & & \ce{^{212}Pb} & $<3.812$ & $<0.01$ \\
     & & \ce{^{212}Bi} & $<3.812$ & $<0.01$ \\
     & & \ce{^{208}Tl} & $<1.370$ & $<0.01$ \\
     & & \ce{^{212}Po} & -- & --\tnote{a} \\
    \multicolumn{5}{l}{Additional External Radiogenic (AER)} \\
     & & \ce{^{234}Th}  & $640 \pm 20$ & $ 1020 \pm 30 $ \\
     & & \ce{^{228}Ra}  & -- 			  & $ <0.01 $ \\
     & & \ce{^{228}Ac}  & $1000 \pm 100$ & $ 365 \pm 7 $ \\
     & & \ce{^{226}Ra}  & $102 \pm 3$ & $ 58 \pm 2 $ \\
     & & \ce{^{214}Bi}  & $340 \pm 30$ & $ 204 \pm 6 $ \\
     & & \ce{^{212}Bi}  & $170 \pm 40$ & $ 260 \pm 10 $ \\
     & & \ce{^{212}Pb}  & $330 \pm 20$ & $ 124 \pm 3 $ \\
     & & \ce{^{210}Pb}  & $84 \pm 3$ & $<0.01$ \\
     & & \ce{^{208}Tl}  & $580 \pm 70$ & $ 236 \pm 5 $ \\
     & & \ce{^{40}K}    & $220 \pm 40$ & $ 260 \pm 10 $ \\
    \multicolumn{5}{l}{Internal Cosmogenic (IC)} \\
     & & \ce{^{181}W}                       &  $39 \pm 2$ & $ 33 \pm 3 $ \\
     & & \ce{^{179}Ta} (M\textsubscript{1}) & $50 \pm 2$ & $ 45 \pm 2 $ \\
     & & \ce{^{179}Ta} (M\textsubscript{2}) & -- & $12 \pm 1$ \\
     & & \ce{^{179}Ta} (M\textsubscript{3}) & -- & $2 \pm 1$ \\
     & & \ce{^{179}Ta} (L\textsubscript{1}) & $136	\pm 2$ & $ 137 \pm 3 $ \\
     & & \ce{^{179}Ta} (L\textsubscript{2}) & $23 \pm 1$ & $ 19 \pm 1 $ \\
     & & \ce{^{179}Ta} (L\textsubscript{3}) & -- & $ <0.01 $ \\
     & & \ce{^{179}Ta} (K)                  & $177 \pm 2$ & $ 172 \pm 3 $ \\
     & & \ce{^{3}H}                         & $24.2 \pm 0.9$ & $ 64 \pm 4 $ \\

\end{longtable}

\begin{tablenotes}
	\footnotesize
	\item[a] Strongly suppressed by branching ratio $\lesssim \SI{0.01}{\percent}$.
	\item[b] $Q$-value of pure alpha decay beyond the upper energy limit of the relevant reference data.
	\item[c] Significant amounts of decays are excluded from the experimental data as pile-up events.
\end{tablenotes}

\end{ThreePartTable}

\end{center}
%\twocolumn
\renewcommand{\arraystretch}{1}

%\end{appendices}

\acknowledgments

This work has been funded by the Deutsche Forschungs\-gemeinschaft (DFG, German Research Foundation) under Germany's Excellence Strategy – EXC 2094 – 390783311 and through the Sonderforschungsbereich (Collaborative Research Center) SFB1258 ‘Neutrinos and Dark Matter in Astro- and Particle Physics’, by the BMBF 05A20WO1 and 05A20VTA and by the Austrian science fund (FWF): I5420-N, W1252-N27. FW was supported through the Austrian research promotion agency (FFG), project ML4CPD. SG was supported through the FWF project STRONG-DM (FG1). JB and HK were funded through the FWF project P 34778-N ELOISE. The Bratislava group acknowledges a partial support provided by the Slovak Research and Development Agency (projects APVV-15-0576 and APVV-21-0377). 
The computational results presented were partially obtained using the CLIP cluster and the Max Planck Computing and Data Facility (MPCDF). 

%\paragraph{Note added.} This is also a good position for notes added
%after the paper has been written.

% Bibliography

%% [A] Recommended: using JHEP.bst file
%% \bibliographystyle{JHEP}
%% \bibliography{biblio.bib}

%% or
%% [B] Manual formatting (see below)
%% (i) We suggest to always provide author, title and journal data or doi:
%% in short all the informations that clearly identify a document.
%% (ii) please avoid comments such as "For a review'', "For some examples",
%% "and references therein" or move them in the text. In general, please leave only references in the bibliography and move all
%% accessory text in footnotes.
%% (iii) Also, please have only one work for each \bibitem.

\bibliographystyle{JHEP}
\bibliography{references.bib}
%\begin{thebibliography}{99}

%\bibitem{a}
%Author,
%\emph{Title},
%\emph{J. Abbrev.} {\bf vol} (year) pg.

%\bibitem{b}
%Author,
%\emph{Title},
%arxiv:1234.5678.

%\bibitem{c}
%Author,
%\emph{Title},
%Publisher (year).

%\end{thebibliography}
\end{document}